\documentclass[a4paper,10pt,journal]{IEEEtran}
\usepackage{amsmath,amsfonts}
\usepackage[caption=false,font=footnotesize]{subfig}
\usepackage{stfloats}
\usepackage{url}
\usepackage{hyperref}
\usepackage{graphicx}
\usepackage{balance}
\usepackage{cite}
\usepackage{mathtools}
\usepackage{booktabs}
\usepackage{amssymb}
\newtheorem{proposition}{Proposition}
\newtheorem{remark}{Remark}

\newtheorem{lemma}{Lemma}
\newtheorem{theorem}{Theorem}
\usepackage{comment}
\newtheorem{corollary}{corollary}
\hypersetup{colorlinks,
	linkcolor=blue,%
	anchorcolor=blue,
    citecolor=blue}

\newcommand{\CN}{\mathcal{CN}}

\begin{document}

\title{\LARGE{Reliable Bistatic Ranging With Unreliable Communication Payloads}}

\author{Fangping Yu, Fan Liu,~\IEEEmembership{Senior Member,~IEEE}, Yifeng Xiong,~\IEEEmembership{Member,~IEEE}, Jie Yang,~\IEEEmembership{Member,~IEEE}, \\Lei Xie,~\IEEEmembership{Member,~IEEE}, and Shi Jin,~\IEEEmembership{Fellow,~IEEE}

\thanks{
This work was supported in part by the National Science and Technology Major Projects of China under Grant 2025ZD1302000, and in part by the National Natural Science Foundation of China (NSFC) under Grant 62522107. \textit{(Corresponding authors: Fan Liu; Yifeng Xiong.)}
}
\thanks{
F. Yu is with the School of Information Science
and Engineering, Southeast University,
Nanjing 210096, China
(e-mail: yu\_fangping@seu.edu.cn).
}
\thanks{
F. Liu is with the National
Mobile Communications Research Laboratory,
Southeast University, Nanjing 210096, China
(e-mail: fan.liu@seu.edu.cn).
}
\thanks{
J. Yang and S. Jin are with the National Mobile Communications Research Laboratory and the Frontiers Science Center for Mobile Information Communication and Security, Southeast University, Nanjing 210096, China (e-mail: yangjie@seu.edu.cn; jinshi@seu.edu.cn).
}
\thanks{
Y. Xiong is with the Department of Communication
Engineering, Beijing University of Posts and
Telecommunications, Beijing 100876, China
(e-mail: yifengxiong@bupt.edu.cn).
}
\thanks{
L. Xie is with the School of Cyber Science
and Engineering, Southeast University,
Nanjing 210096, China
(e-mail: leixie@seu.edu.cn).
}
}



\maketitle

\begin{abstract}
This paper addresses a key challenge in bistatic integrated sensing and communications (ISAC): \textit{How to achieve reliable ranging using communication data signals that cannot be reliably recovered?} In modern communication systems, data payloads typically occupy far more time-frequency resources than pilots, offering substantially more observations to improve sensing accuracy. Exploiting them requires recovering the transmitted symbols to form sensing references. However, propagation and reflection losses can severely weaken target echoes. The sensing receiver may therefore be unable to recover data transmitted at a high rate selected for communication users. We first show that, at low signal-to-noise ratios (SNRs), using hard symbol decisions as sensing references can introduce a persistent bias in the estimated channel impulse response (CIR). This bias produces an error floor and spurious peaks in the power delay profile (PDP) that persist even after coherent accumulation over a large number of observations. To address this limitation, we develop a pilot-initialized expectation-maximization (EM) estimator that iteratively refines the CIR using posterior symbol moments. Under local regularity conditions and successful initialization, we establish that the EM method converges to the local maximum-likelihood (ML) solution for a sufficiently large number of OFDM data symbols. The converged estimator is asymptotically unbiased, where the CIR estimation mean-square error (MSE) at each delay bin decays as $O(1/M)$ by accumulating $M$ data symbols. Thus, estimation error can vanish through sufficient accumulation even while symbol recovery remains highly unreliable. Numerical results validate the proposed method, demonstrating substantially higher target detection probabilities than pilot-only and hard-decision-based bistatic ranging under the same low-SNR conditions.
\end{abstract}


\section{Introduction}
\label{sec:introduction}

\IEEEPARstart{I}{ntegrated} sensing and communications (ISAC) is envisioned as a key capability of 6G networks, which integrates information transmission and environmental sensing through shared spectrum and hardware~\cite{ref1,ref2,ref3,ref4,ref5,ref6}. By reusing ubiquitous wireless signals, ISAC turns communication infrastructure into a source of environmental measurements. These measurements can supply physical artificial intelligence (AI) with information about object locations, motion, and environmental changes, supporting perception and decision-making in intelligent transportation, robotics, and low-altitude applications~\cite{ref2,ref7,ref8,ref9}.

Directly reusing communication signals for sensing, commonly referred to as \textit{communication-centric ISAC}, offers a practical route to deployment by building on standardized waveforms and existing network infrastructure~\cite{ref10,ref11,ref12,ref13,ref14}. The known pilot symbols embedded in these signals provide readily available sensing references, but occupy only a small fraction of each transmission frame. Relying solely on pilots therefore leaves most resources carrying data payloads unused for sensing. For a given observation interval, this limits the echo energy available for accumulation and consequently restricts sensing accuracy and coverage. To exploit the majority of resources, recent studies have investigated sensing with communication data signals \cite{LiuIceberg2025,Lu2024RandomPrecoding,ref21,ref24,ref25,KeskinPayload2026}. By analyzing the statistical properties of their correlation and ambiguity functions, these works have revealed how modulation, constellation design, and pulse shaping affect sidelobes and ranging performance \cite{LiuIceberg2025}.

Most existing studies assume a monostatic ISAC architecture, where the transmitter (Tx) and sensing receiver (Rx) are located at the same node \cite{ref10,ref11,LiuIceberg2025}. Both pilots and transmitted data symbols are therefore known to the sensing processor and can be directly used as reference signals. However, receiving target echoes during transmission requires full-duplex operation and strong suppression of leakage from the local Tx \cite{ref15,ref16,ref17,ref19,ref20}. In cellular deployments, interference from neighboring base stations further complicates echo reception. These difficulties motivate bistatic sensing, where the Tx and sensing Rx are spatially separated \cite{ref21,ref22,Valiulahi2026LANM}. A receive-only sensing node avoids local Tx leakage and the associated self-interference cancellation burden, simplifying Rx hardware and signal processing~\cite{ref21,ref22}. Bistatic operation also supports coordinated sensing across multiple network nodes. By forming different Tx--Rx pairs, the network can observe the same environment from complementary geometries, exploit spatial diversity, and extend sensing coverage beyond that of a single monostatic node~\cite{ref2,ref4}.

\begin{figure}[t]
    \centering
    \includegraphics[width=0.9\columnwidth]{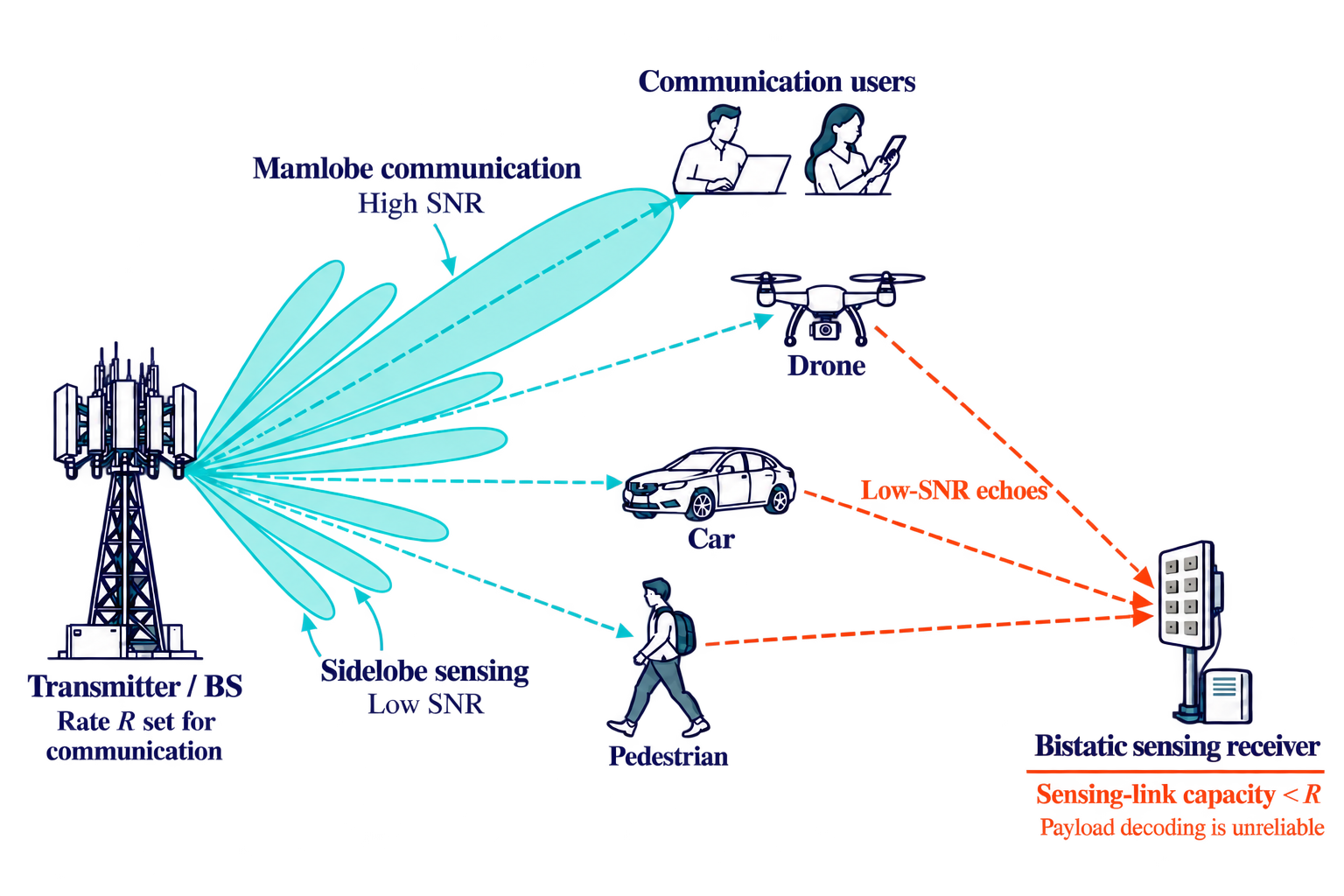}
    \caption{Illustration of the low-SNR challenge in bistatic sensing.}
    \label{fig:bistatic_scenario}
\end{figure}

The separated Tx--Rx topology, however, imposes a critical challenge in exploiting communication data for sensing. Unlike a monostatic node, the bistatic sensing Rx does not know the transmitted data symbols. Conventional processing therefore first decodes the data and leverages the recovered symbols as sensing references. This approach requires sufficient received SNR, which is not always available on the bistatic sensing link. For example, Fig.~\ref{fig:bistatic_scenario} illustrates a base station serving communication users through its antenna mainlobe, while targets outside the mainlobe are illuminated by sidelobes. The lower sidelobe gain, together with propagation and target reflection losses, can make the echoes at the sensing Rx much weaker than the signals received by communication users. More severely, the higher SNR at these users allows the Tx to select a high data rate, which may exceed what the weaker sensing link can support. Under a Gaussian channel model, this mismatch can be expressed as
\begin{equation*}
    \underbrace{\log\!\left(1+\mathrm{SNR}_{\mathrm{sen}}\right)}_{C_{\mathrm{sen}}}
    < R
    \leq
    \underbrace{\log\!\left(1+\mathrm{SNR}_{\mathrm{com}}\right)}_{C_{\mathrm{com}}},
    \label{eq:rate_capacity_mismatch}
\end{equation*}
where $R$ is the transmission rate, and $C_{\mathrm{sen}}$ and $C_{\mathrm{com}}$ are the sensing- and communication-link capacities, respectively. In this regime, reliable payload recovery at the sensing Rx is not possible at the prescribed rate, even with perfect channel state information (CSI). Yet the received data signals still carry rich information about the targets. This raises a central question: \textit{How to achieve reliable bistatic sensing when the communication data cannot be reliably recovered?}

Existing studies have developed several approaches to exploiting unknown data payloads for bistatic sensing. A common approach uses hard symbol decisions or reconstructs a symbol sequence from decoded bits, then treats the resulting symbols as known sensing references \cite{ref21,ref25}. Following the latter approach, Brunner et al. \cite{ref21} reconstructed the transmitted frame through communication decoding and used it for sensing, demonstrating the benefits of full-frame processing and examining the effects of decoding failures. Henninger et al. \cite{ref25} improved data recovery by allocating lower-order modulation symbols as pseudo-pilots, trading a small reduction in communication spectral efficiency for better sensing performance. Alternatively, Keskin et al. \cite{ref24} proposed an iterative Rx that uses linear minimum mean-square error (LMMSE)-based soft symbol estimates to refine the channel estimate, improving multi-target detection and reducing the gap to the perfect-data benchmark. More explicit treatments of payload uncertainty have also been investigated. Keskin et al. \cite{KeskinPayload2026} analyzed sidelobe degradation under an additive symbol-error model and developed an uncertainty-aware minimum mean-square error Rx. Xie et al. \cite{Xie2025} instead exploited the statistics of unknown Gaussian payloads without recovering individual symbols, deriving a generalized likelihood ratio test and large-sample approximations of its detection performance. These studies demonstrate the performance gain of exploiting unknown or uncertain payloads for sensing. However, the achievable sensing performance when data recovery remains unreliable at low SNR is still not fully understood.

The accumulation behavior is particularly important in the low-SNR regime, since weak echoes require observations over multiple transmissions to achieve accurate sensing. In monostatic ISAC, our previous work \cite{LiuIceberg2025} established an ``iceberg-in-the-sea'' decomposition of the average squared auto-correlation function (ACF). With random but known transmitted symbols, coherent averaging over $M$ independent transmissions reduces the mean-square fluctuations (the ``sea level'') caused by random data by a factor of $M$, while preserving the deterministic response (the ``iceberg''). In bistatic ISAC, processing gains have also been discussed, and existing studies have related sidelobe performance to observation length or compared multi-slot processing with known and unknown symbols \cite{Xie2025,Hatami2026,ref24}. However, these analyses do not establish the scaling behavior for the sensing performance when finite-alphabet data symbols remain unreliably recovered at low SNR. In particular, hard-decision processing treats each selected constellation symbol as the true transmitted reference, discarding the uncertainty in that decision. Whether this approximation introduces an estimation bias and limits the accumulation gain remains to be established. Understanding these limitations is an important step towards developing a bistatic ISAC Rx that accounts for symbol uncertainty and achieves increasingly accurate sensing as more observations are accumulated.

To address these challenges, we develop an EM-based technique for bistatic ranging with unreliable communication payloads. EM provides a general framework to incorporate unknown data into channel estimation through their posterior statistics \cite{ref29,ref30,Ladaycia2019,RednerWalker1984,Ma2004EM,Balakrishnan2017EM}. Here, pilots and uncertain payloads are jointly exploited to estimate the channel impulse response (CIR) for target detection, with particular attention to sensing performance at low SNR and the gains achievable through accumulation. The main contributions are summarized as follows:
\begin{itemize}
    \item We characterize the limitations of hard-decision-based bistatic sensing. We derive the CIR estimation error caused by imperfect symbol references, and show that hard decisions introduce bias-dependent floor and persistent spurious peaks in the power delay profile (PDP) that remain even after accumulating over a large number of observations. This identifies a limitation of accumulation with hard-decision references and motivates processing that explicitly accounts for symbol uncertainty.

    \item We develop a pilot-initialized EM Rx for bistatic ranging with unreliable payloads. The Rx jointly uses pilots and posterior symbol moments to iteratively refine the CIR, while incorporating its finite delay support throughout the estimation process. Target detection is then performed using the resulting PDP. This enables communication payloads to contribute to sensing even when individual data symbols cannot be reliably recovered.

    \item We establish the local convergence and asymptotic unbiasedness of the proposed Rx. Under local regularity conditions and successful pilot initialization, EM converges to the corresponding local maximum-likelihood estimator (MLE) for sufficient number of payload observations. We further prove strong consistency of this estimator and conditional asymptotic unbiasedness of the converged EM output, showing that the CIR estimation bias vanishes as more payload observations are accumulated despite unreliable symbol recovery.

    \item We derive the asymptotic accumulation gain of the EM Rx. Under
    the same conditions, with fixed pilot resources and CIR dimension, the
    CIR estimation MSE at each delay bin satisfies
    \[
        \operatorname{MSE}
        =
        \frac{c}{M}
        +
        o(M^{-1}),
    \]
    where $M$ is the number of payload OFDM symbols and $c$ is determined
    by the per-symbol observed Fisher information. Thus, despite unreliable
    data recovery, payload accumulation reduces the CIR estimation error
    power at the rate of $1/M$.
\end{itemize}
Numerical results further validate the effectiveness of the EM-based bistatic ranging method, demonstrating substantial suppression of spurious range peaks and improved target detection probabilities compared with pilot-only and hard-decision-based sensing.

The remainder of this paper is organized as follows.
Section~\ref{sec:system_model} presents the bistatic OFDM signal model
and analyzes the persistent CIR bias caused by hard-decision-based
estimation. Section~\ref{sec:em_bistatic_sensing} develops the
pilot-initialized EM estimator and establishes its local convergence,
asymptotic unbiasedness, and per-delay-bin MSE scaling.
Section~\ref{sec:numerical_results} presents the numerical results, and
Section~\ref{sec:conclusion} concludes the paper.

\subsection*{Notations}

Scalars are denoted by italic letters, vectors and matrices by bold
letters, and sets, events, and operators by calligraphic letters.
The superscripts $(\cdot)^T$, $(\cdot)^H$, and $(\cdot)^*$ denote
transpose, Hermitian transpose, and complex conjugation, respectively.
The operators $\operatorname{Re}(\cdot)$, $\operatorname{Im}(\cdot)$,
$|\cdot|$, $\|\cdot\|$, $\operatorname{tr}(\cdot)$, and
$\operatorname{Diag}(\cdot)$ have their standard meanings, and
$\odot$ denotes elementwise multiplication. The operators
$\mathbb{E}[\cdot]$, $\operatorname{Var}(\cdot)$, and $\Pr(\cdot)$
denote expectation, variance, and probability. The distributions
$\CN(\boldsymbol{\mu},\mathbf{C})$ and
$\mathcal{N}(\mu,\sigma^2)$ denote complex and real Gaussian
distributions, respectively. For Hermitian matrices,
$\succeq$ and $\succ$ denote positive-semidefinite and
positive-definite ordering. The symbols $\nabla$, $\nabla^2$, and $D$
denote the gradient, Hessian, and Jacobian, respectively. The notations
$O(\cdot)$ and $o(\cdot)$ have their standard asymptotic
meanings, and ``a.s.'' denotes almost-sure convergence.

\section{Hard-Decision-Based Bistatic Ranging}
\label{sec:system_model}

\subsection{System Model}

Consider an OFDM-based bistatic sensing system consisting of a Tx and a spatially separated sensing Rx. The Tx illuminates the targets while serving communication users, and the Rx collects the target echoes propagating along the Tx--target--Rx paths. Within each sensing interval, the transmitted frame contains one pilot OFDM symbol followed by $M$ data OFDM symbols, each occupying $N$ subcarriers. The bistatic channel is assumed to remain unchanged over these $M+1$ symbols.

The CIR is supported on the first $L$ delay samples, where $1\leq L\leq N$, and the cyclic prefix is sufficiently long to cover this delay support. The CIR vector is denoted by $\mathbf{g}=[g_0,\ldots,g_{L-1}]^T\in\mathbb{C}^{L}$. Suppose that $K_{\rm tar}$ resolvable targets are present. Under an on-grid delay model, the $q$th target has a complex coefficient $\alpha_q$ and a delay index $k_q\in\{0,1,\ldots,L-1\}$. The CIR is then given by
\begin{equation}
    g_k
    =
    \sum_{q=1}^{K_{\rm tar}}
    \alpha_q\delta[k-k_q],
    \qquad
    k=0,1,\ldots,L-1,
    \label{eq:sparse_channel}
\end{equation}
where $\delta[\cdot]$ denotes the Kronecker delta.

Let $\mathbf{F}\in\mathbb{C}^{N\times N}$ denote the unitary discrete Fourier transform (DFT) matrix, and let $\mathbf{F}_L\in\mathbb{C}^{N\times L}$ contain its first $L$ columns. After cyclic-prefix removal and an $N$-point DFT, the received pilot observation is
\begin{equation}
    \mathbf{y}_p
    = \mathbf{p}\odot\mathbf{h} + \mathbf{n}=
    \mathbf{p}\odot\mathbf{F}_L\mathbf{g}
    +
    \mathbf{n},
    \label{eq:pilot_model}
\end{equation}
where $\mathbf{h}:= \mathbf{F}_L\mathbf{g}$ denotes the frequency-domain channel vector, $\mathbf{p}\in\mathbb{C}^{N}$ is the known pilot vector, and $\mathbf{n}\sim\CN(\mathbf{0},\sigma_p^2\mathbf{I}_N)$. The pilot symbol has constant modulus such that $|p_n|^2=1$.

The observation corresponding to the $m$th data OFDM symbol is
\begin{equation}
    \mathbf{y}_m
    = \mathbf{x}_m\odot\mathbf{h} + \mathbf{z}_m=
    \mathbf{x}_m\odot\mathbf{F}_L\mathbf{g}
    +
    \mathbf{z}_m,
    \quad
    m=1,\ldots,M,
    \label{eq:data_model}
\end{equation}
where $\mathbf{x}_m\in\mathbb{C}^{N}$ contains the transmitted data symbols and $\mathbf{z}_m\sim\CN(\mathbf{0},\sigma^2\mathbf{I}_N)$. Unlike $\mathbf{p}$, the data vector $\mathbf{x}_m$ is random unknown to the bistatic sensing Rx. Its entries are independently and uniformly drawn from a finite constellation $\mathcal{A}$ and normalized such that $\mathbb{E}\left[|x_{m,n}|^2\right]=1$. The data vectors are independent across OFDM symbols, while the noise vectors are independent of all transmitted symbols.

\subsection{CIR Estimation for Ranging}

Under the on-grid model in \eqref{eq:sparse_channel}, each target contributes a non-zero CIR coefficient at its corresponding delay bin. Ranging and target detection are therefore performed using the PDP of the estimated CIR, where $|\widehat{g}_k|^2$ is the received power at the $k$th delay bin. Peaks in the PDP indicate candidate target delays, and their amplitudes can be compared with a detection threshold. The sensing performance thus depends directly on the accuracy of the CIR estimate.

Consider first monostatic sensing, where the sensing processor has direct access to all transmitted pilot and data symbols. Their observations can be jointly used to estimate the CIR through solving the following least-squares (LS) problem:
\begin{equation}
\begin{aligned}
\min_{\mathbf{g}\in\mathbb{C}^{L}}\;
\frac{1}{\sigma_p^2}
\left\|
\mathbf{y}_p
-
\mathbf{p}\odot\mathbf{F}_L\mathbf{g}
\right\|^2
+
\frac{1}{\sigma^2}
\sum_{m=1}^{M}
\left\|
\mathbf{y}_m
-
\mathbf{x}_m\odot\mathbf{F}_L\mathbf{g}
\right\|^2 .
\end{aligned}
\label{eq:general_wls}
\end{equation}
The resulting weighted LS estimate is
\begin{equation}
\begin{aligned}
\widehat{\mathbf{g}}
={}&
\Bigg[
\mathbf{F}_L^H
\left(
\frac{1}{\sigma_p^2}\mathbf{I}_N
+
\frac{1}{\sigma^2}
\sum_{m=1}^{M}
\operatorname{Diag}(|\mathbf{x}_m|^2)
\right)
\mathbf{F}_L
\Bigg]^{-1}
\\
&\cdot
\mathbf{F}_L^H
\left(
\frac{\mathbf{p}^*\odot\mathbf{y}_p}{\sigma_p^2}
+
\frac{1}{\sigma^2}
\sum_{m=1}^{M}
\mathbf{x}_m^*\odot\mathbf{y}_m
\right).
\end{aligned}
\label{eq:general_channel_estimate}
\end{equation}
The matrix inside the inverse is lower bounded by $\sigma_p^{-2}\mathbf{I}_L$ and is therefore positive definite. Hence, \eqref{eq:general_channel_estimate} is uniquely defined.

In bistatic sensing, the sensing Rx does not know the transmitted payload symbols $\mathbf{x}_m$. Conventional reference-based processing therefore first estimates these symbols from the received observations and then substitutes their estimates for $\mathbf{x}_m$ in \eqref{eq:general_wls}. For example, hard-decision processing uses symbol estimates $\widehat{\mathbf{x}}_m\in\mathcal{A}^{N}$ and obtains the CIR estimate by replacing $\mathbf{x}_m$ with $\widehat{\mathbf{x}}_m$ in \eqref{eq:general_channel_estimate}. When symbol recovery is unreliable, these estimated references may significantly differ from the transmitted symbols and consequently distort the estimated CIR and its PDP. The effect of hard-decision errors on CIR estimation and coherent accumulation is analyzed next.

\subsection{Bias of Hard-Decision-Based CIR Estimation}

Hard-decision processing replaces the unknown payload symbols
$\mathbf{x}_m$ in \eqref{eq:general_wls} with their decisions
$\widehat{\mathbf{x}}_m\in\mathcal{A}^{N}$. Let
$\widehat{\mathbf{g}}$ denote the resulting CIR estimate. Its bias is
defined as
\begin{equation}
\mathbf{b}
:=
\mathbb{E}
\left[
\widehat{\mathbf{g}}
\right]
-
\mathbf{g},
\label{eq:decision_bias}
\end{equation}
where the expectation is taken over the payload symbols and Rx
noise.

To identify the sources of this bias, define
\begin{equation}
\mathbf{W}
:=
\frac{1}{\sigma_p^2}\mathbf{I}_N
+
\frac{1}{\sigma^2}
\sum_{m=1}^{M}
\operatorname{Diag}
\left(
|\widehat{\mathbf{x}}_m|^2
\right).
\label{eq:decision_W}
\end{equation}
Substituting the pilot and data models in \eqref{eq:pilot_model} and
\eqref{eq:data_model} into the CIR estimator gives
\begin{equation}
\widehat{\mathbf{g}}-\mathbf{g}
=
\left(
\mathbf{F}_L^H\mathbf{W}\mathbf{F}_L
\right)^{-1}
\mathbf{F}_L^H\mathbf{q},
\label{eq:decision_error}
\end{equation}
where
\begin{align}
\mathbf{q}
={}&
\underbrace{
\frac{1}{\sigma_p^2}
\mathbf{p}^*\odot\mathbf{n}
}_{\text{pilot-weighted noise}}
\nonumber+
\underbrace{
\frac{1}{\sigma^2}
\sum_{m=1}^{M}
(\mathbf{F}_L\mathbf{g})\odot
\left(
\mathbf{x}_m\odot\widehat{\mathbf{x}}_m^*
-
|\widehat{\mathbf{x}}_m|^2
\right)
}_{\text{reference mismatch}}\\
&+
\underbrace{
\frac{1}{\sigma^2}
\sum_{m=1}^{M}
\mathbf{z}_m\odot\widehat{\mathbf{x}}_m^*
}_{\text{decision-weighted noise}}.
\label{eq:decision_q}
\end{align}
It follows from \eqref{eq:decision_error} that
\begin{equation*}
\mathbf{b}
=
\mathbb{E}
\left[
\left(
\mathbf{F}_L^H\mathbf{W}\mathbf{F}_L
\right)^{-1}
\mathbf{F}_L^H\mathbf{q}
\right].
\end{equation*}

With exact symbol knowledge,
$\widehat{\mathbf{x}}_m=\mathbf{x}_m$, and the reference-mismatch term
vanishes. Moreover, $\mathbf{W}$ depends only on the transmitted
symbols and is independent of the noise. Since both the
pilot-weighted noise and the data noise
$\mathbf{z}_m\odot\mathbf{x}_m^*$ have zero conditional mean, the
resulting CIR estimate is unbiased. This gives the known-payload
benchmark, as in monostatic sensing.

With hard decisions, symbol errors make the reference-mismatch term
non-zero. Such errors occur frequently in the low-SNR regime and make
this mismatch more pronounced. In addition,
$\widehat{\mathbf{x}}_m$ depends on the same Rx noise
$\mathbf{z}_m$ appearing in the observation, so the
decision-weighted noise generally has a non-zero mean. The dependence
of $\mathbf{W}$ on the noisy hard decisions further contributes to
the bias in \eqref{eq:decision_bias}.

The effect on sensing can be seen from the expected PDP. At the $k$th delay
bin,
\begin{equation}
\mathbb{E}
\left[
|\widehat g_k|^2
\right]
=
|g_k+b_k|^2+\operatorname{Var}(\widehat g_k).
\label{eq:decision_power}
\end{equation}
At a non-target delay bin, where $g_k=0$, this reduces to
$|b_k|^2+\operatorname{Var}(\widehat g_k)$. Accumulating independent payload observations
suppresses the zero-mean random fluctuations, whereas a nonvanishing
bias remains through $|b_k|^2$. Its variation across delay bins
can raise the PDP floor, create spurious range peaks, and degrade
target detection. We next use BPSK as a case study to show how
hard-decision processing produces such a persistent CIR bias.

\subsection{Case Study: Bistatic Ranging With Hard-Detected BPSK}

Consider equiprobable normalized BPSK payloads,
$x_{m,n}\in\{-1,+1\}$, and let
$\mathbf{h}=\mathbf{F}_L\mathbf{g}$ denote the frequency-domain
channel. The pilot observation first provides the initial CIR estimate
and its frequency-domain response:
\begin{equation}
\widehat{\mathbf{g}}_0
:=
\mathbf{F}_L^H
\left(
\mathbf{p}^*\odot\mathbf{y}_p
\right),
\qquad
\widehat{\mathbf{h}}_0
=
\mathbf{F}_L\widehat{\mathbf{g}}_0.
\label{eq:bpsk_initial_estimate}
\end{equation}
The latter is used to equalize the payload symbols before hard
decisions are made.

Define the phase error of the initial channel estimate as
\begin{equation}
\psi_n
:=
\arg(h_n)-\arg(\widehat h_{0,n}),
\qquad
n=0,\ldots,N-1.
\label{eq:bpsk_phase}
\end{equation}
We consider a successful pilot initialization satisfying
$h_n\neq0$, $\widehat h_{0,n}\neq0$, and
$|\psi_n|<\pi/2$ on all subcarriers. The last condition only excludes
a phase reversal and is satisfied when the pilot SNR is sufficient to
identify the correct channel half-plane. The resulting hard decision
after equalization is
\begin{equation}
\widehat{x}_{m,n}
=
\operatorname{sign}
\left(
\operatorname{Re}
\left\{
\frac{y_{m,n}}{\widehat h_{0,n}}
\right\}
\right).
\label{eq:bpsk_decision}
\end{equation}
Since the pilot and BPSK decisions have unit modulus, the resulting
CIR estimate in \eqref{eq:general_channel_estimate} reduces to
\begin{equation}
\begin{aligned}
\widehat{\mathbf{g}}
={}&
\frac{\sigma^2}
{\sigma^2+M\sigma_p^2}
\widehat{\mathbf{g}}_0
+
\frac{M\sigma_p^2}
{\sigma^2+M\sigma_p^2}
\mathbf{F}_L^H
\left(
\frac{1}{M}
\sum_{m=1}^{M}
\mathbf{y}_m\odot\widehat{\mathbf{x}}_m^*
\right).
\end{aligned}
\label{eq:bpsk_hard_cir_estimate}
\end{equation}
Its conditional bias is characterized below.

\begin{figure}[t]
    \centering
    \includegraphics[width=0.9\columnwidth]{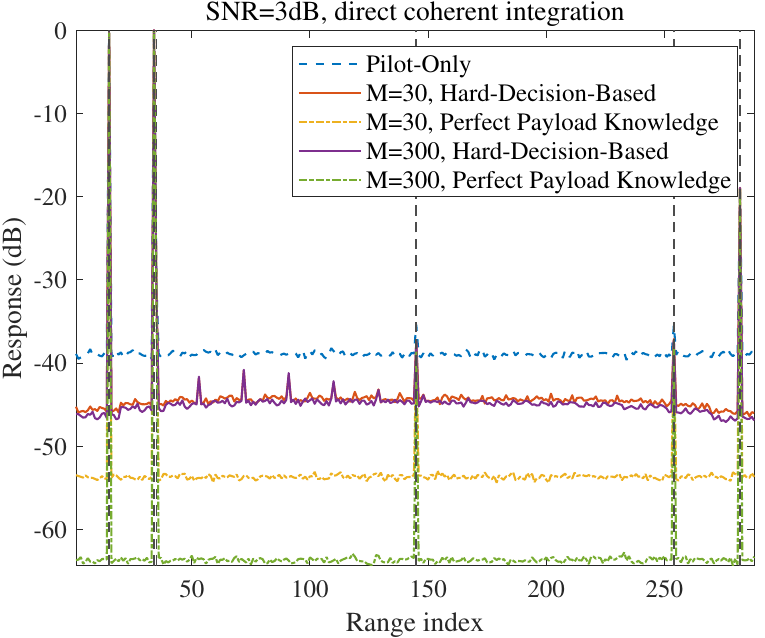}
    \caption{Illustration of the impact of decision-induced bias on the
    PDP for different numbers of payload observations at
    $\mathrm{SNR}=3$ dB.}
    \label{fig:bias_impact_direct_coherent}
\end{figure}

\begin{proposition}[Persistent BPSK hard-decision bias]
\label{prop:bpsk_bias}
Conditioned on the pilot initialization described above, the CIR
estimation bias based on hard-detected BPSK symbols is
\begin{equation}
\begin{aligned}
\mathbf{b}
\left(
\widehat{\mathbf{h}}_0
\right)
&=
\frac{\sigma^2}
{\sigma^2+M\sigma_p^2}
\left(
\widehat{\mathbf{g}}_0-\mathbf{g}
\right)
+
\frac{M\sigma_p^2}
{\sigma^2+M\sigma_p^2}
\mathbf{F}_L^H\boldsymbol{\xi},
\end{aligned}
\label{eq:bpsk_bias}
\end{equation}
where
$\boldsymbol{\xi}=[\xi_0,\ldots,\xi_{N-1}]^T$, with
\begin{equation}
\xi_n
:=
\mathbb{E}
\left[
y_{m,n}\widehat{x}_{m,n}^*
\mid
\widehat{\mathbf{h}}_0
\right]
-
h_n
\label{eq:bpsk_xin}
\end{equation}
denoting the conditional mean error contributed by one hard-detected
payload symbol on subcarrier $n$. Moreover,
\begin{equation}
\lim_{M\to\infty}
\mathbf{b}
\left(
\widehat{\mathbf{h}}_0
\right)
=
\mathbf{F}_L^H\boldsymbol{\xi}
\neq
\mathbf{0}.
\label{eq:bpsk_bias_limit}
\end{equation}
\end{proposition}

\begin{IEEEproof}
See Appendix~\ref{app:bpsk_bias}.
\end{IEEEproof}

\begin{remark}
Equation~\eqref{eq:bpsk_bias} shows that the contribution of the pilot
initialization error vanishes as $M$ increases, whereas the
hard-decision residual converges to the non-zero limit
$\mathbf{F}_L^H\boldsymbol{\xi}$. This bias-limited accumulation is
illustrated in Fig.~\ref{fig:bias_impact_direct_coherent}, where
$\mathrm{SNR}=1/\sigma^2=3$~dB and the uncoded BPSK error rate is on
the order of $10^{-2}$ under unit channel gain and accurate phase
initialization. With only one pilot symbol, the initial estimate has a
high noise floor that masks the weak targets. Using $M=30$
hard-detected payloads lowers the PDP floor relative to the pilot-only
estimate, but increasing $M$ from $30$ to $300$ provides little further
improvement. In contrast, the perfect-payload-knowledge benchmark gains
approximately $10$~dB and clearly reveals the weak targets. Thus,
decision errors produce a persistent CIR bias that appears as an
elevated PDP floor and spurious range peaks, consistent with
Proposition~\ref{prop:bpsk_bias}. To recover the accumulation gain with
unreliable payloads, the next section develops an EM estimator that
replaces hard decisions with posterior symbol moments.
\end{remark}

\section{EM-Based Bistatic Ranging}
\label{sec:em_bistatic_sensing}

The main difficulty in exploiting data-bearing OFDM symbols is that
their payload is unknown at the sensing Rx. Direct joint
estimation of the CIR $\mathbf{g}$ and the $MN$ discrete symbols may
require examining up to $|\mathcal{A}|^{MN}$ symbol configurations.
Hard decisions avoid this combinatorial search but, as shown above, can introduce a systematic CIR bias that cannot be suppressed by accumulation.

The EM algorithm avoids both exhaustive search and premature hard
decisions by treating the payload symbols as latent variables in the
standard incomplete-data maximum-likelihood formulation
\cite{ref29,ref30}. At each iteration, the E-step computes their
posterior moments conditioned on the current CIR estimate, and the
M-step updates the $L$-tap CIR using these soft statistics. The detailed procedure is presented below.

\subsection{EM-Based CIR Estimator}

Let $\mathcal{X}:=\{\mathbf{x}_1,\ldots,\mathbf{x}_M\}$ denote the set of unknown payload symbols, and $\mathcal{Y}:=\{\mathbf{y}_1,\ldots,\mathbf{y}_M\}$ the set of observed data signals. Define
\begin{equation}
    \mathcal{D}
    =
    \{\mathbf{y}_p\}
    \cup
    \mathcal{Y}
    \cup
    \mathcal{X}.
\end{equation}
With independent Gaussian noise and equiprobable payload symbols, the complete data
log-likelihood is
\begin{equation}
\begin{aligned}
\ell_c(\mathbf{g};\mathcal{D})
&=
-\frac{1}{\sigma_p^2}
\left\|
\mathbf{y}_p-\mathbf{p}\odot\mathbf{F}_L\mathbf{g}
\right\|^2
\\
&\quad
-\frac{1}{\sigma^2}
\sum_{m=1}^{M}
\left\|
\mathbf{y}_m-\mathbf{x}_m\odot\mathbf{F}_L\mathbf{g}
\right\|^2
+c_0,
\end{aligned}
\label{eq:complete_log_likelihood}
\end{equation}
where $c_0$ is independent of $\mathbf{g}$.

\subsubsection{E-step}

Given the current CIR estimate $\widehat{\mathbf{g}}^{(t)}$, let
\begin{equation}
\widehat{\mathbf{h}}^{(t)}
=
\mathbf{F}_L\widehat{\mathbf{g}}^{(t)}
\end{equation}
and define the EM auxiliary function as
\begin{equation}
\mathcal{L}_M(\mathbf{g}\left.\right|\hat{\mathbf{g}}^{(t)})
:=
\mathbb{E}_{\mathcal{X}\mid
\mathcal{Y},\widehat{\mathbf{g}}^{(t)}}
\left[
\ell_c(\mathbf{g};\mathcal{D})
\right].
\label{eq:em_auxiliary}
\end{equation}
The posterior distribution factorizes across payload symbols. For
$a\in\mathcal{A}$, the posterior probability of $x_{m,n}=a$ is
\begin{equation}
\gamma_{m,n}^{(t)}(a)
=
\frac{
\exp\left(
-\left|y_{m,n}-\widehat h_n^{(t)}a\right|^2/\sigma^2
\right)
}{
\displaystyle
\sum_{b\in\mathcal{A}}
\exp\left(
-\left|y_{m,n}-\widehat h_n^{(t)}b\right|^2/\sigma^2
\right)
}.
\label{eq:posterior_probability}
\end{equation}
The corresponding posterior first- and second-order moments are
\begin{align}\label{eq:posterior_moments}
\overline{x}_{m,n}^{(t)}
=
\sum_{a\in\mathcal{A}}
a\,\gamma_{m,n}^{(t)}(a),\qquad
\overline{|x_{m,n}|^2}^{\,(t)}
=
\sum_{a\in\mathcal{A}}
|a|^2\gamma_{m,n}^{(t)}(a).
\end{align}
Collect these moments into
$\overline{\mathbf{x}}_m^{(t)}$ and
$\overline{|\mathbf{x}_m|^2}^{\,(t)}$, respectively, and define
\begin{equation}
\mathbf{W}^{(t)}
:=
\frac{1}{\sigma_p^2}\mathbf{I}_N
+
\frac{1}{\sigma^2}
\sum_{m=1}^{M}
\operatorname{Diag}
\left(
\overline{|\mathbf{x}_m|^2}^{\,(t)}
\right),
\label{eq:em_W}
\end{equation}
and
\begin{equation}
\mathbf{r}^{(t)}
:=
\frac{\mathbf{p}^*\odot\mathbf{y}_p}{\sigma_p^2}
+
\frac{1}{\sigma^2}
\sum_{m=1}^{M}
\left(
\overline{\mathbf{x}}_m^{(t)}
\right)^*
\odot\mathbf{y}_m.
\label{eq:em_r}
\end{equation}
The auxiliary function then reduces to
\begin{equation}
\begin{aligned}
\mathcal{L}_M(\mathbf{g}\left.\right|\hat{\mathbf{g}}^{(t)})
&=
-\mathbf{g}^H
\mathbf{F}_L^H\mathbf{W}^{(t)}\mathbf{F}_L
\mathbf{g}
+
2\operatorname{Re}
\left\{
\mathbf{g}^H\mathbf{F}_L^H\mathbf{r}^{(t)}
\right\}
+c,
\end{aligned}
\label{eq:em_auxiliary_quadratic}
\end{equation}
where $c$ is independent of $\mathbf{g}$.

\subsubsection{M-step}
With the posterior moments from the E-step held fixed, maximize
$\mathcal{L}_M(\mathbf{g}\left.\right|\hat{\mathbf{g}}^{(t)})$ with respect to $\mathbf{g}$. This gives
\begin{equation}
    \widehat{\mathbf{g}}^{(t+1)}
    =
    \left(
        \mathbf{F}_L^H
        \mathbf{W}^{(t)}
        \mathbf{F}_L
    \right)^{-1}
    \mathbf{F}_L^H
    \mathbf{r}^{(t)}.
    \label{eq:finite_support_M_step}
\end{equation}
The algorithm is initialized using the pilot observation as
\begin{equation}
    \widehat{\mathbf{g}}^{(0)}
    =
    \mathbf{F}_L^H
    \left(
        \mathbf{p}^*\odot\mathbf{y}_p
    \right),
    \label{eq:finite_support_init}
\end{equation}
The E- and M-steps are repeated until convergence or until a
prescribed number of iterations is reached.
\begin{remark}
The vector $\mathbf{r}^{(t)}$ can be interpreted as the output of
matched filtering with symbol references. The known pilot is used
directly, whereas each payload observation is matched to the
posterior mean $\overline{\mathbf{x}}_m^{(t)}$, allowing unreliable
symbols to be naturally downweighted.
\end{remark}

\subsection{Asymptotic Unbiasedness of the EM Estimator}
\label{subsec:em_asymptotic_unbiasedness}

To show that EM avoids the persistent hard-decision bias as the
payload length increases, we establish its local asymptotic
unbiasedness. The key step is to show that, under a suitable
initialization, EM converges to the local observed data MLE,
after which standard ML asymptotics can be applied.

For the likelihood analysis, let $\mathbf{u}\in\mathbb{C}^{L}$ denote
a generic channel parameter and define its real representation as
\begin{equation}
\boldsymbol{\theta}(\mathbf{u})
:=
\left[
\operatorname{Re}\{\mathbf{u}\}^{T},
\operatorname{Im}\{\mathbf{u}\}^{T}
\right]^{T}
\in\mathbb{R}^{2L}.
\end{equation}
The true parameter is
$\boldsymbol{\theta}^{\star}:=\boldsymbol{\theta}(\mathbf{g})$.
For a generic payload observation $\mathbf{y}$, define the
observed data score at the true channel as
\begin{equation}
\mathbf{s}(\mathbf{y};\boldsymbol{\theta}^{\star})
:=
\left.
\nabla_{\boldsymbol{\theta}}
\log p_{\boldsymbol{\theta}}(\mathbf{y})
\right|_{\boldsymbol{\theta}=\boldsymbol{\theta}^{\star}}.
\end{equation}
The corresponding per-symbol observed Fisher information is
\begin{equation}
\mathbf{J}_o
:=
\mathbb{E}_{\boldsymbol{\theta}^{\star}}
\left[
\mathbf{s}(\mathbf{y};\boldsymbol{\theta}^{\star})
\mathbf{s}(\mathbf{y};\boldsymbol{\theta}^{\star})^T
\right].
\label{eq:observed_fisher}
\end{equation}
The true channel is called regular if
$\mathbf{J}_o\succ\mathbf{0}$.

The observed data log-likelihood based on the pilot and $M$ payload
symbols is
\begin{equation}
\begin{aligned}
\ell_{o,M}(\mathbf{u})
&:=
\log
p_{\mathbf{u}}
\left(
\mathbf{y}_p,\mathcal{Y}\mid\mathbf{p}
\right)
=\log
\sum_{\mathcal{X}}
\exp
\left\{
\ell_c(\mathbf{u};\mathcal{D})
\right\},
\end{aligned}
\label{eq:observed_log_likelihood}
\end{equation}
where the sum is over all possible payload collections
$\mathcal{X}$.

\begin{lemma}[Convergence of EM to the local MLE]
\label{prop:em_to_mle}
Suppose that the true channel $\mathbf{g}$ is regular. There exist a
deterministic compact neighborhood $\Theta$ of $\mathbf{g}$ and a
deterministic open neighborhood
$\mathcal{U}\subset\Theta$ containing $\mathbf{g}$ such that, with
probability one, for all sufficiently large $M$, the maximizer of
$\ell_{o,M}$ over $\Theta$ is unique and lies in
$\operatorname{int}(\Theta)$. Denote this maximizer by
$\widehat{\mathbf{g}}_M^{\rm ML}$. If
$\widehat{\mathbf{g}}^{(0)}\in\mathcal{U}$, then
\begin{equation}
\lim_{t\rightarrow\infty}
\widehat{\mathbf{g}}_M^{(t)}
=
\widehat{\mathbf{g}}_M^{\rm ML}.
\label{eq:em_to_mle}
\end{equation}
\end{lemma}

\begin{IEEEproof}
See Appendix~\ref{app:em_to_mle}.
\end{IEEEproof}
Lemma~\ref{prop:em_to_mle} identifies a
neighborhood $\mathcal{U}$ independent of $M$ such that, for all sufficiently large $M$,
the EM iterates converge to the unique local MLE whenever
the pilot-based initialization lies in $\mathcal{U}$. The asymptotic
properties of the EM estimator can therefore be obtained from those
of the local MLE.

\begin{proposition}[Strong consistency of the local MLE]
\label{prop:mle_consistency}
The local MLE satisfies
\begin{equation}
\lim_{M\rightarrow\infty}
\widehat{\mathbf{g}}_M^{\rm ML}
=
\mathbf{g},
\qquad
\text{a.s.}
\label{eq:mle_consistency}
\end{equation}
\end{proposition}

\begin{IEEEproof}
See Appendix~\ref{app:mle_consistency}.
\end{IEEEproof}
Together with Lemma~\ref{prop:em_to_mle},
Proposition~\ref{prop:mle_consistency} shows that, under the stated
initialization, the EM estimator converges almost surely to the true
channel as the number of payload OFDM symbols increases.

Define the successful initialization event as
\begin{equation}
\mathcal{E}
:=
\left\{
\widehat{\mathbf{g}}^{(0)}
\in
\mathcal{U}
\right\}.
\label{eq:initialization_event}
\end{equation}
From \eqref{eq:finite_support_init}, we have $\widehat{\mathbf{g}}^{(0)}-\mathbf{g}
\sim
\mathcal{CN}
\left(
\mathbf{0},
\sigma_p^2\mathbf{I}_L
\right)$. Since $\mathcal{U}$ is an open neighborhood of $\mathbf{g}$, it follows that
$\operatorname{Pr}(\mathcal{E})>0$. To make the estimator well defined for every finite $M$, define $\widehat{\mathbf{g}}_M^{(\infty)}$ as the ordinary EM limit whenever
the limit exists and belongs to $\Theta$, and set it to
$\widehat{\mathbf{g}}^{(0)}$ otherwise. This completion only covers
finite-sample cases for which local EM convergence is not guaranteed.
By Lemma~\ref{prop:em_to_mle}, conditioned on $\mathcal{E}$,
$\widehat{\mathbf{g}}_M^{(\infty)}$ agrees almost surely with the
ordinary EM limit and the local MLE for all sufficiently
large $M$.

\begin{theorem}[Local asymptotic unbiasedness of the EM estimator]
\label{thm:em_asymptotic_unbiasedness}
At a regular channel point,
\begin{equation}
\lim_{M\rightarrow\infty}
\mathbb{E}
\left[
\widehat{\mathbf{g}}_M^{(\infty)}
\mid
\mathcal{E}
\right]
=
\mathbf{g}.
\label{eq:em_asymptotic_unbiasedness}
\end{equation}
\end{theorem}

\begin{IEEEproof}
See Appendix~\ref{app:em_asymptotic_unbiasedness}.
\end{IEEEproof}
\begin{remark}
Theorem~\ref{thm:em_asymptotic_unbiasedness} shows that an
asymptotically unbiased CIR estimate does not require every payload
symbol to be detected correctly. Under the stated initialization, the
EM iterations converge to the local MLE, which approaches the
true CIR as $M$ increases. Thus, retaining payload uncertainty through
soft posterior information avoids the persistent bias caused by hard
decisions.
\end{remark}

\subsection{Asymptotic MSE of the EM Estimator}
\label{subsec:em_mse}

Asymptotic unbiasedness alone does not quantify how fast the
estimation error decreases as more payload symbols are accumulated.
We therefore characterize the per-delay-bin MSE of the converged EM
estimator.

\begin{theorem}[Asymptotic MSE of the EM estimator]
\label{thm:em_mse}
At a regular channel point, conditioned on $\mathcal{E}$, for each
fixed $k=0,1,\ldots,L-1$,
\begin{equation}
\begin{aligned}
&
\mathbb{E}
\left[
\left|
\widehat g_{M,k}^{(\infty)}
-
g_k
\right|^2
\middle|
\mathcal{E}
\right]
\\
&\quad=
\frac{
[\mathbf{J}_o^{-1}]_{k+1,k+1}
+
[\mathbf{J}_o^{-1}]_{L+k+1,L+k+1}
}{M}
+
o(M^{-1}).
\end{aligned}
\label{eq:em_mse_result}
\end{equation}
\end{theorem}

\begin{IEEEproof}
See Appendix~\ref{app:em_mse}.
\end{IEEEproof}

Theorem~\ref{thm:em_mse} shows that the MSE at each
delay bin decreases asymptotically as $1/M$. At a non-target bin with
$g_k=0$, this MSE is also the expected PDP level, which
therefore follows the same scaling. Importantly, this $1/M$ scaling remains valid at low SNR under the stated regularity and initialization conditions, even when individual
payload symbols cannot be decoded reliably. By using posterior soft
information, EM lowers the non-target PDP floor as $M$ increases, while
the target responses converge to their true values.

\section{Numerical Results}
\label{sec:numerical_results}

\begin{table}[t]
\caption{Simulation Parameters}
\label{tab:sim_parameters}
\centering
\footnotesize
\setlength{\tabcolsep}{4pt}
\renewcommand{\arraystretch}{1.05}
\begin{tabular}{p{0.46\columnwidth} p{0.34\columnwidth}}
\toprule
\textbf{Parameter} & \textbf{Value / Description} \\
\midrule
Modulation format (data) & 16-QAM \\
Modulation format (pilot) & QPSK \\
Number of subcarriers & $N=\text{3276}$ \\
Number of pilot symbols & $1$ \\
Data OFDM symbols (default) & $M=\text{13}$ \\
Delay search length & $L=\text{288}$ \\
Channel model & sparse multipath \\
\bottomrule
\end{tabular}
\end{table}

In this section, simulations are presented to validate the preceding
analysis and evaluate the proposed EM-based estimator. In particular,
we examine the impact of hard-decision bias and the coherent
accumulation gain obtained by increasing the number of payload OFDM
symbols. The pilot and payload symbols are assumed to have the same
SNR. Unless otherwise stated, $\mathrm{SNR}:=1/\sigma^2$ is set
to $3$~dB, for which the uncoded 16-QAM symbol error rate is on the
order of $10^{-1}$. The remaining default parameters are listed in
Table~\ref{tab:sim_parameters}.

\begin{figure}[t]
    \centering
    \includegraphics[width=0.9\linewidth]{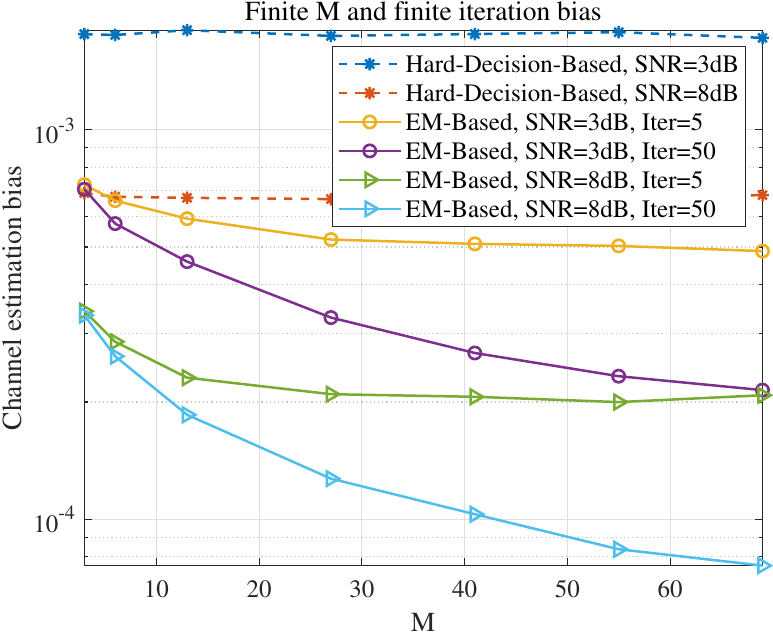}
    \caption{CIR estimation bias versus the number of data OFDM symbols $M$ for hard-decision-based and EM methods at different SNRs and EM iteration counts.}
    \label{fig:finite_M_iter}
\end{figure}

Fig.~\ref{fig:finite_M_iter} examines how the number of data
OFDM symbols and the EM iteration count affect the empirical
CIR estimation bias. At both SNRs, the hard-decision-based
curves remain nearly flat as $M$ increases, indicating that
additional observations do not remove the systematic bias
introduced by symbol decisions. The EM-based curves, by
comparison, decrease with $M$, and a larger iteration count
generally yields a lower bias at the same SNR. For both fixed iteration counts, however, the decrease
gradually slows as $M$ grows. This leveling is more apparent
with fewer iterations. With more iterations, the bias continues
to decrease over a wider range of $M$ before its rate of
reduction also becomes smaller. The two trends have different origins: additional EM iterations move the estimate closer to its local ML solution, whereas increasing $M$ reduces the statistical error of that solution. 

\begin{figure}[t]
    \centering

    \subfloat[Range sidelobe profiles under hard-decision-based and EM-based estimation.]
    {
        \centering
        \includegraphics[width=0.9\linewidth]{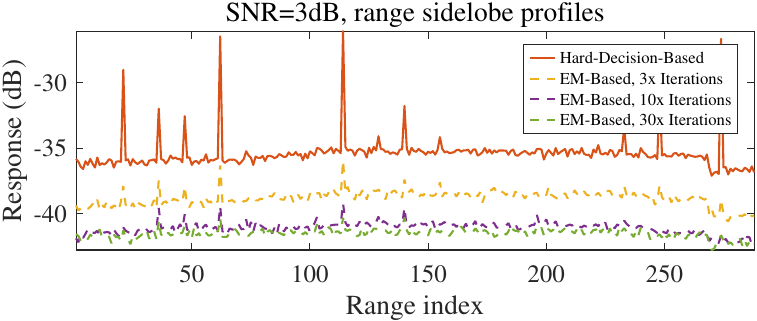}
        \label{fig:range_sidelobe_profile}
    }

    \vspace{0.5em}

    \subfloat[Empirical cumulative distribution function (ECDF) of sidelobe levels at $\mathrm{SNR}=3\,\mathrm{dB}$.]
    {
        \centering
        \includegraphics[width=0.9\linewidth]{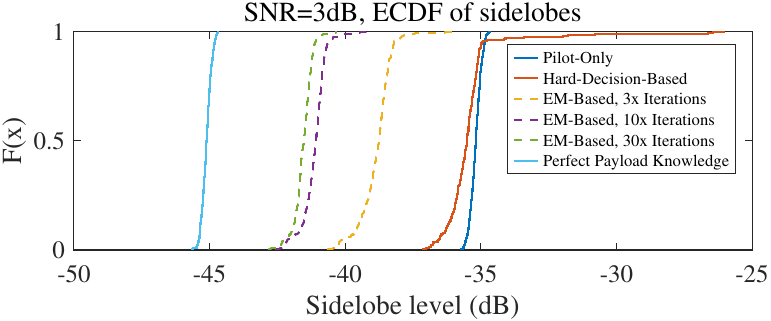}
        \label{fig:ecdf_sidelobe_5dB}
    }
    
    \caption{PDP sidelobe suppression performance of hard-decision-based and EM-based CIR estimation methods.}
    \label{fig:sidelobe_suppression_results}
\end{figure}


To evaluate the iterative behavior of the EM-based estimator, Fig.~\ref{fig:sidelobe_suppression_results} compares the PDP sidelobe profiles and their empirical cumulative distribution functions (ECDFs) under different iteration numbers. As shown in Fig.~\ref{fig:range_sidelobe_profile}, the hard-decision-based method produces pronounced sidelobe peaks in non-target range bins, whereas EM progressively suppresses these peaks. Substantial improvement is achieved within the first few iterations, with further sidelobe reduction after 10 iterations and a smoother, lower sidelobe profile after 30 iterations. Consistently, Fig.~\ref{fig:ecdf_sidelobe_5dB} shows a progressive leftward shift of the ECDF curves, indicating reduced sidelobe levels. These results demonstrate that EM achieves effective sidelobe suppression with relatively few iterations, while additional iterations provide further refinement.

\begin{figure}[t]
    \centering
    \includegraphics[width=0.9\linewidth]{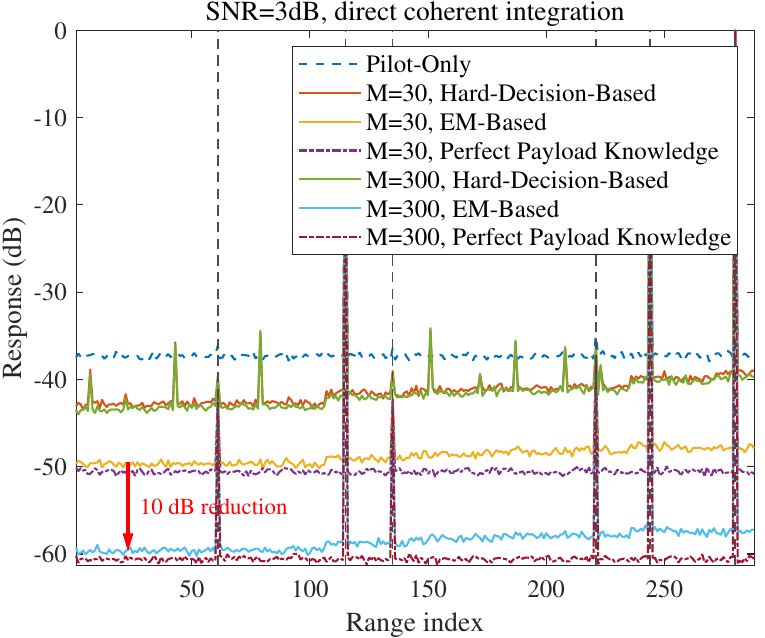}
    \caption{PDPs obtained from pilot-only, hard-decision-based, EM-based, and perfect-payload-knowledge-aided CIR estimates at $\mathrm{SNR}=3\,\mathrm{dB}$ for $M=30$ and $M=300$.}
    \label{fig:iceberg_sea}
\end{figure}

Fig.~\ref{fig:iceberg_sea} compares the PDPs at an SNR of $3\,\mathrm{dB}$. The dominant target peaks occur at the same range indices for all four methods. In the non-target region, the hard-decision-based method exhibits an elevated background and several distinct spurious peaks, whereas the EM-based PDP approaches the perfect-payload-knowledge reference. When $M$ increases from $30$ to $300$, the EM non-target background decreases by approximately $10\,\mathrm{dB}$, while the hard-decision-based background changes little and its spurious peaks remain visible. At non-target delay bins, this trend agrees with \eqref{eq:em_mse_result}. Increasing $M$ tenfold predicts an asymptotic $10\,\mathrm{dB}$ reduction in the estimation error power. The persistent hard-decision-based PDP floor is consistent with a systematic decision-induced bias that is not removed by increasing the number of payload observations.


\begin{figure}[!t]
    \centering
    \subfloat[Hard-decision-based method.]{
        \includegraphics[width=0.9\columnwidth]{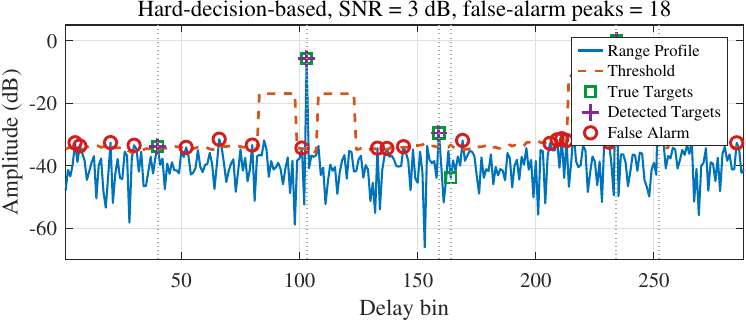}
        \label{fig:cfar_profile_hard}
    }

    \vspace{1mm}

    \subfloat[EM-based method.]{
        \includegraphics[width=0.9\columnwidth]{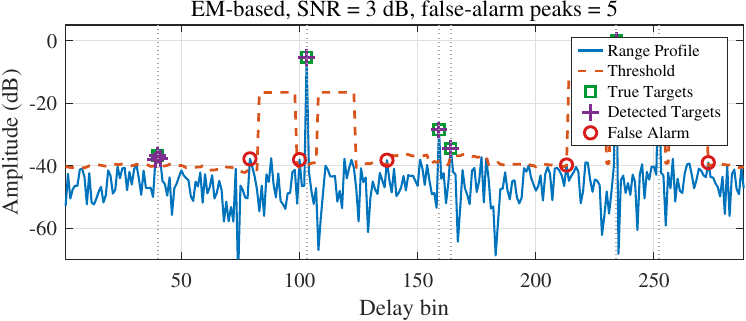}
        \label{fig:cfar_profile_em}
    }

    \caption{PDPs and CA-CFAR detection results of the
    hard-decision-based and EM-based methods at
    \(\mathrm{SNR}=3\) dB.}
    \label{fig:CFAR_thre}
\end{figure}

Fig.~\ref{fig:CFAR_thre} compares the PDPs and cell-averaging constant false alarm rate (CA-CFAR) detection results of the hard-decision-based and EM-based methods at \(\mathrm{SNR}=3\) dB. The red circles indicate false alarms at non-target delay bins whose PDP values exceed the CFAR threshold. The decision-induced bias in the hard-decision-based method elevates several non-target PDP values, resulting in false alarms. In contrast, the EM-based method suppresses these spurious peaks and produces fewer false alarms under the same detection setting.

\begin{figure}[t]
    \centering
    \includegraphics[width=0.9\linewidth]{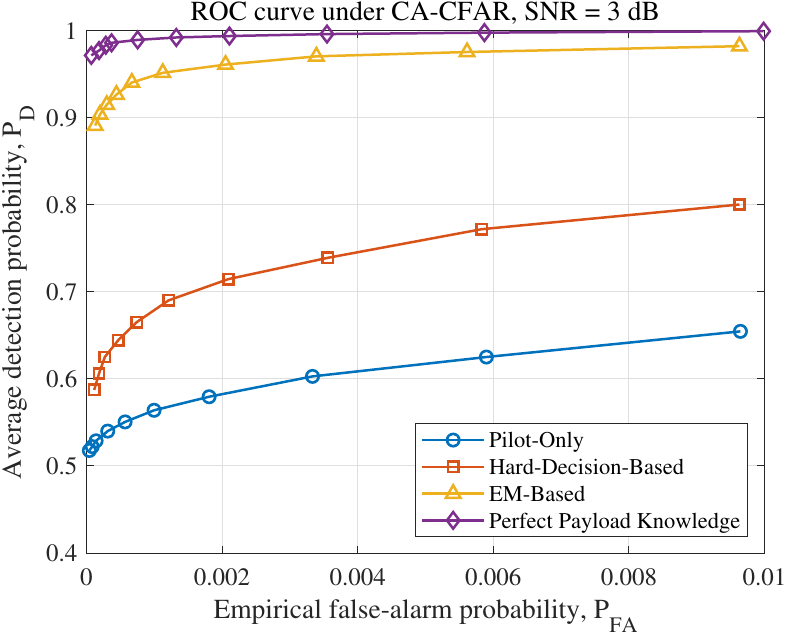}
    \caption{ROC curves under CA-CFAR detection at
    \(\mathrm{SNR}=3\) dB.}
    \label{fig:CFAR_ROC}
\end{figure}

Fig.~\ref{fig:CFAR_ROC} further evaluates the detection performance through receiver operating characteristic (ROC) curves under CA-CFAR detection. The perfect-payload-knowledge matched filter achieves the best performance, while the EM-based method closely approaches this benchmark and outperforms the hard-decision-based method over a wide range of false-alarm probabilities. The pilot-only method exhibits the lowest detection probability due to its limited use of observations. These results demonstrate that exploiting posterior soft information mitigates decision errors and improves the detection--false-alarm tradeoff.

\section{Conclusion}
\label{sec:conclusion}

This paper studied bistatic ranging with communication payloads that
cannot be reliably recovered at the sensing Rx. We showed that,
at low SNR, hard symbol decisions introduce a persistent CIR bias,
producing an error floor and spurious peaks in the PDP that cannot be
removed through coherent accumulation. To address this problem, we
developed a pilot-initialized EM estimator that refines the CIR using
posterior symbol moments. Under local regularity conditions and
successful initialization, the EM iterations converge to the local
ML solution for sufficiently many payload symbols. The converged
estimator is asymptotically unbiased, with its MSE at each delay bin
decaying as $O(1/M)$. Simulations with 16-QAM payloads confirmed that
EM can accumulate sensing information even when symbol recovery remains
unreliable, substantially improving CA-CFAR target detection over
pilot-only and hard-decision-based ranging.

\appendices

\section{Proof of Proposition~\ref{prop:bpsk_bias}}
\label{app:bpsk_bias}

Taking the conditional expectation of
\eqref{eq:bpsk_hard_cir_estimate} gives
\begin{equation}
\begin{aligned}
\mathbb E
\left[
\widehat{\mathbf g}
\mid
\widehat{\mathbf h}_0
\right]
-\mathbf g
&=
\frac{\sigma^2}
{\sigma^2+M\sigma_p^2}
\left(
\widehat{\mathbf g}_0-\mathbf g
\right)
\\
&+
\frac{M\sigma_p^2}
{\sigma^2+M\sigma_p^2}
\mathbf F_L^H
\left(
\mathbb E
\left[
\mathbf y_m\odot\widehat{\mathbf x}_m^*
\mid
\widehat{\mathbf h}_0
\right]
-\mathbf h
\right).
\end{aligned}
\label{eq:app_bpsk_cir_bias}
\end{equation}
The vector in parentheses is $\boldsymbol{\xi}$ defined in
\eqref{eq:bpsk_xin}. Its $n$th entry is
\begin{equation}
\begin{aligned}
\xi_n
&=
\mathbb{E}
\left[
y_{m,n}\widehat x_{m,n}^*
\mid
\widehat{\mathbf{h}}_0
\right]
-h_n
\\
&=
h_n
\left(
\mathbb{E}
\left[
x_{m,n}\widehat x_{m,n}
\mid
\widehat{\mathbf{h}}_0
\right]
-1
\right)
+
\mathbb{E}
\left[
z_{m,n}\widehat x_{m,n}
\mid
\widehat{\mathbf{h}}_0
\right].
\end{aligned}
\label{eq:app_bpsk_residual_decomposition}
\end{equation}

Using the phase error $\psi_n$, \eqref{eq:bpsk_decision} becomes
\begin{equation}
\widehat x_{m,n}
=
\operatorname{sign}
\left(
|h_n|x_{m,n}\cos\psi_n
+
\operatorname{Re}
\left\{
\frac{\widehat h_{0,n}^*}{|\widehat h_{0,n}|}z_{m,n}
\right\}
\right).
\end{equation}
Since $x_{m,n}^2=1$, multiplying the argument of the sign function
by $x_{m,n}$ gives
\begin{equation}
|h_n|\cos\psi_n
+
x_{m,n}
\operatorname{Re}
\left\{
\frac{\widehat h_{0,n}^*}{|\widehat h_{0,n}|}z_{m,n}
\right\}
\sim
\mathcal{N}
\left(
|h_n|\cos\psi_n,
\frac{\sigma^2}{2}
\right).
\end{equation}
Define
\begin{equation}
\rho_n
:=
\frac{\sqrt{2}|h_n|\cos\psi_n}{\sigma}>0.
\label{eq:app_bpsk_t}
\end{equation}
Let $\phi(t):=(2\pi)^{-1/2}e^{-t^2/2}$ and
$Q(t):=\int_t^\infty\phi(u)\,du$. The symbol error probability is
$Q(\rho_n)$, and hence
\begin{equation}
\mathbb{E}
\left[
x_{m,n}\widehat x_{m,n}
\mid
\widehat{\mathbf{h}}_0
\right]
=
1-2Q(\rho_n).
\label{eq:app_bpsk_decision_moment}
\end{equation}

Gaussian symmetry gives
\begin{equation}
\begin{aligned}
&
\mathbb{E}
\left[
\operatorname{Re}
\left\{
\frac{\widehat h_{0,n}^*}{|\widehat h_{0,n}|}z_{m,n}
\right\}
\widehat x_{m,n}
\left.\right|
\widehat{\mathbf{h}}_0
\right]
\\
&=
2
\int_{-|h_n|\cos\psi_n}^{\infty}
\frac{u}{\sqrt{\pi}\sigma}
\exp
\left(
-\frac{u^2}{\sigma^2}
\right)
\,du
=
\sqrt{2}\sigma\phi(\rho_n).
\end{aligned}
\end{equation}
The imaginary part of the rotated noise is independent of the
decision and has zero mean. Rotating back gives
\begin{equation}
\mathbb{E}
\left[
z_{m,n}\widehat x_{m,n}
\mid
\widehat{\mathbf{h}}_0
\right]
=
\sqrt{2}\sigma\phi(\rho_n)
\frac{\widehat h_{0,n}}{|\widehat h_{0,n}|}.
\label{eq:app_bpsk_noise_moment}
\end{equation}

Using \eqref{eq:app_bpsk_decision_moment} and
\eqref{eq:app_bpsk_noise_moment} in
\eqref{eq:app_bpsk_residual_decomposition} yields
\begin{equation}
\xi_n
=
-2h_nQ(\rho_n)
+
\sqrt{2}\sigma\phi(\rho_n)
\frac{\widehat h_{0,n}}{|\widehat h_{0,n}|}.
\label{eq:app_bpsk_xi_closed_form}
\end{equation}
From \eqref{eq:app_bpsk_xi_closed_form},
\begin{align}
\operatorname{Re}
\left\{
\widehat h_{0,n}^*\xi_n
\right\}
&\nonumber=
\sqrt{2}\sigma
|\widehat h_{0,n}|
\left[
\phi(\rho_n)-\rho_nQ(\rho_n)
\right]\\
&=
\sqrt{2}\sigma
|\widehat h_{0,n}|
\int_{\rho_n}^\infty
(u-\rho_n)\phi(u)\,du
>0.
\end{align}
Since
$\widehat{\mathbf{h}}_0
=\mathbf{F}_L\widehat{\mathbf{g}}_0$,
\begin{equation}
0
<
\operatorname{Re}
\left\{
\widehat{\mathbf{h}}_0^H\boldsymbol{\xi}
\right\}
=
\operatorname{Re}
\left\{
\widehat{\mathbf{g}}_0^H
\mathbf{F}_L^H\boldsymbol{\xi}
\right\}.
\end{equation}
Therefore,
$\mathbf{F}_L^H\boldsymbol{\xi}\neq\mathbf{0}$. Taking
$M\to\infty$ in \eqref{eq:bpsk_bias} proves
\eqref{eq:bpsk_bias_limit}.

\section{Proof of Lemma~\ref{prop:em_to_mle}}
\label{app:em_to_mle}

All derivatives below are taken with respect to the $2L$ real
coordinates defined in
Section~\ref{subsec:em_asymptotic_unbiasedness}. The proof first
establishes local regularity of the likelihood, then shows that the
population EM map is locally contractive. This contraction is
subsequently transferred to the sample EM map, whose unique fixed
point is finally identified with the local MLE.

\subsubsection{Local Regularity}
For a generic payload observation $\mathbf{y}$,
\begin{equation}
p_{\mathbf{u}}(\mathbf{y})
=
\prod_{n=0}^{N-1}
\frac{1}{|\mathcal{A}|\pi\sigma^2}
\sum_{a\in\mathcal{A}}
\exp
\left(
-\frac{
|y_n-[\mathbf{F}_L\mathbf{u}]_na|^2
}{\sigma^2}
\right).
\end{equation}
Since $[\mathbf{F}_L\mathbf{u}]_n$ is affine in the real coordinates of
$\mathbf{u}$ and $\mathcal{A}$ is finite, differentiating the Gaussian
mixture likelihood produces finite sums of polynomial residual terms
weighted by posterior probabilities. Hence, for $k=0,1,2,3$,
\begin{equation}
\left\|
D^k\log p_{\mathbf{u}}(\mathbf{y})
\right\|
\le
C_k
\left(
1+\|\mathbf{y}\|^{q_k}
\right),
\label{eq:likelihood_envelope}
\end{equation}
uniformly for $\mathbf{u}$ in any fixed compact neighborhood of
$\mathbf{g}$, where
$D^0\log p_{\mathbf{u}}(\mathbf{y})
\equiv\log p_{\mathbf{u}}(\mathbf{y})$,
$C_k<\infty$, and $q_k<\infty$.

Define the posterior moments
\begin{align}
\overline{\mathbf{x}}(\mathbf{u};\mathbf{y})
:=
\mathbb{E}_{\mathbf{u}}
\left[
\mathbf{x}\mid\mathbf{y}
\right],\qquad
\mathbf{d}(\mathbf{u};\mathbf{y})
:=
\mathbb{E}_{\mathbf{u}}
\left[
|\mathbf{x}|^2\mid\mathbf{y}
\right].
\end{align}
The same argument gives, for $k=0,1,2$,
\begin{equation}
\begin{aligned}
&
\left\|
D^k\overline{\mathbf{x}}(\mathbf{u};\mathbf{y})
\right\|
+
\left\|
D^k\mathbf{d}(\mathbf{u};\mathbf{y})
\right\|
\le
C'_k
\left(
1+\|\mathbf{y}\|^{q'_k}
\right),
\end{aligned}
\label{eq:posterior_envelope}
\end{equation}
uniformly on every fixed compact neighborhood of $\mathbf{g}$.
The true payload distribution is a finite Gaussian mixture and
therefore has moments of every finite order. The bounds in
\eqref{eq:likelihood_envelope} and
\eqref{eq:posterior_envelope} are thus integrable, which justifies the
differentiation under the expectation and the uniform strong laws used
below.

Define the population payload log-likelihood as
\begin{equation}
\overline{\ell}_o(\mathbf{u})
:=
\mathbb{E}_{\mathbf{g}}
\left[
\log p_{\mathbf{u}}(\mathbf{y})
\right].
\end{equation}
At the true channel,
\begin{equation}
\nabla\overline{\ell}_o(\mathbf{g})
=
\mathbf{0},
\qquad
\nabla^2\overline{\ell}_o(\mathbf{g})
=
-\mathbf{J}_o.
\end{equation}
Since
$\nabla^2\overline{\ell}_o(\mathbf{g})=-\mathbf{J}_o\prec\mathbf{0}$
and the Hessian is continuous, it remains uniformly negative definite
in a sufficiently small neighborhood of $\mathbf{g}$. Specifically,
there exist $R>0$ and $\lambda>0$ such that, for
\begin{equation}
\Theta
:=
\left\{
\mathbf{u}\in\mathbb{C}^{L}:
\|\mathbf{u}-\mathbf{g}\|\le R
\right\},
\end{equation}
we have
\begin{equation}
\nabla^2\overline{\ell}_o(\mathbf{u})
\preceq
-2\lambda\mathbf{I}_{2L},
\qquad
\mathbf{u}\in\Theta.
\label{eq:population_concavity}
\end{equation}
Thus, $\overline{\ell}_o$ is strongly concave on $\Theta$. Since
$\nabla\overline{\ell}_o(\mathbf{g})=\mathbf{0}$, the true channel
$\mathbf{g}$ is its unique maximizer over $\Theta$.

\subsubsection{Local Contraction of the Population EM Map}

The E-step computes the posterior moments separately for each payload
observation, while the M-step averages these moments over the $M$
observations to update the channel. To characterize the large-$M$
behavior, we replace the empirical averages appearing in the M-step
with their expectations under the true channel $\mathbf{g}$. The pilot
contribution vanishes after normalization by $M$, yielding the
deterministic population EM map
\begin{equation}
\mathcal{T}(\mathbf{u})
:=
\mathbf{A}(\mathbf{u})^{-1}
\boldsymbol{\beta}(\mathbf{u}),
\label{eq:population_em_operator}
\end{equation}
where
\begin{equation}
\begin{aligned}
\mathbf{A}(\mathbf{u})
&=
\frac{1}{\sigma^2}
\mathbf{F}_L^H
\mathbb{E}_{\mathbf{g}}
\left[
\operatorname{Diag}
\big(
\mathbf{d}(\mathbf{u};\mathbf{y})
\big)
\right]
\mathbf{F}_L,
\\
\boldsymbol{\beta}(\mathbf{u})
&=
\frac{1}{\sigma^2}
\mathbf{F}_L^H
\mathbb{E}_{\mathbf{g}}
\left[
\overline{\mathbf{x}}(\mathbf{u};\mathbf{y})^*
\odot\mathbf{y}
\right].
\end{aligned}
\end{equation}
Here, $\mathbf{u}$ is the current channel value used to compute the
posterior moments, whereas the outer expectation is taken under the
true channel $\mathbf{g}$. The population map is used to analyze
the large-$M$ behavior of the sample EM iterations.

At $\mathbf{u}=\mathbf{g}$, the posterior distribution is evaluated
under the true model. The law of iterated expectation gives
\begin{equation}
\begin{aligned}
\mathbb{E}_{\mathbf{g}}
\left[
\mathbf{d}(\mathbf{g};\mathbf{y})
\right]
&=
\mathbb{E}_{\mathbf{g}}
\left[
|\mathbf{x}|^2
\right]
=
\mathbf{1}_N,
\\
\mathbb{E}_{\mathbf{g}}
\left[
\overline{\mathbf{x}}(\mathbf{g};\mathbf{y})^*
\odot\mathbf{y}
\right]
&=
\mathbb{E}_{\mathbf{g}}
\left[
\mathbf{x}^*\odot\mathbf{y}
\right]
=
\mathbf{F}_L\mathbf{g}.
\end{aligned}
\end{equation}
Using $\mathbf{F}_L^H\mathbf{F}_L=\mathbf{I}_L$, we obtain
\begin{equation}
\mathbf{A}(\mathbf{g})
=
\frac{1}{\sigma^2}\mathbf{I}_L,
\qquad
\boldsymbol{\beta}(\mathbf{g})
=
\frac{1}{\sigma^2}\mathbf{g}.
\end{equation}
Therefore,
\begin{equation}
\mathcal{T}(\mathbf{g})
=
\mathbf{g},
\label{eq:population_fixed_point}
\end{equation}
so the true channel is a fixed point of the population EM map.

To determine whether this fixed point is locally attractive, consider
the complete data model in which both $\mathbf{x}$ and $\mathbf{y}$ are
observed. For one payload OFDM symbol, its log-likelihood is
\begin{equation}
\ell_c^{(1)}
\left(
\mathbf{u};\mathbf{x},\mathbf{y}
\right)
=
-\frac{1}{\sigma^2}
\left\|
\mathbf{y}
-
\mathbf{x}\odot\mathbf{F}_L\mathbf{u}
\right\|^2
+c,
\end{equation}
where $c$ is independent of $\mathbf{u}$. By leveraging the fact $\mathbb{E}
\left[
\operatorname{Diag}
\left(
|\mathbf{x}|^2
\right)
\right]
=
\mathbf{I}_N$, the corresponding
complete data Fisher information at the true channel is
\begin{equation}
\mathbf{J}_c
:=
-
\mathbb{E}_{\mathbf{g}}
\left[
\left.
\nabla_{\boldsymbol{\theta}}^2
\ell_c^{(1)}
\left(
\mathbf{u};\mathbf{x},\mathbf{y}
\right)
\right|_{\mathbf{u}=\mathbf{g}}
\right]=
\frac{2}{\sigma^2}
\mathbf{I}_{2L}.
\end{equation}

By the EM rate identity and the missing-information decomposition
\cite[Sec.~3, Theorem~4]{ref29},
\begin{equation}
D\mathcal{T}(\mathbf{g})
=
\mathbf{J}_c^{-1}
\left(
\mathbf{J}_c-\mathbf{J}_o
\right)
=
\mathbf{I}_{2L}
-
\mathbf{J}_c^{-1}\mathbf{J}_o.
\label{eq:em_rate_matrix}
\end{equation}
Regularity and the missing-information principle give
$\mathbf{0}\prec\mathbf{J}_o\preceq\mathbf{J}_c$. Since
$\mathbf{J}_c=(2/\sigma^2)\mathbf{I}_{2L}$, all eigenvalues of
$\mathbf{J}_c^{-1}\mathbf{J}_o$ lie in $(0,1]$. Therefore,
\begin{equation}
\left\|
D\mathcal{T}(\mathbf{g})
\right\|_2
=
\max_i
\left|
1-
\lambda_i
\left(
\mathbf{J}_c^{-1}\mathbf{J}_o
\right)
\right|
<1.
\label{eq:population_jacobian_contraction}
\end{equation}

By continuity of $D\mathcal{T}$, there exist $r\in(0,R)$ and
$\kappa\in(0,1)$ such that, on the closed neighborhood
\begin{equation}
\mathcal{C}
:=
\left\{
\mathbf{u}\in\mathbb{C}^L:
\|\mathbf{u}-\mathbf{g}\|\le r
\right\},
\end{equation}
we have
\begin{equation}
\sup_{\mathbf{u}\in\mathcal{C}}
\left\|
D\mathcal{T}(\mathbf{u})
\right\|_2
\le
\kappa.
\label{eq:population_contraction}
\end{equation}
Consequently, for every $\mathbf{u}\in\mathcal{C}$,
\begin{equation}
\left\|
\mathcal{T}(\mathbf{u})-\mathbf{g}
\right\|
\le
\kappa
\left\|
\mathbf{u}-\mathbf{g}
\right\|
\le
\kappa r
<r.
\label{eq:population_mapping}
\end{equation}
Thus, $\mathcal{T}$ is a contraction on $\mathcal{C}$ and maps
$\mathcal{C}$ into its interior, which is defined as
\begin{equation}
\mathcal{U}
:=
\operatorname{int}(\mathcal{C})
=
\left\{
\mathbf{u}\in\mathbb{C}^L:
\|\mathbf{u}-\mathbf{g}\|<r
\right\}.
\end{equation}

\subsubsection{Contraction of the Sample EM Map}

We next show that the sample EM map approaches the population map as
$M$ increases. Dividing both terms in the M-step by $M$ does not change
the update, which can be written as
\begin{equation}
\mathcal{T}_M(\mathbf{u})
=
\mathbf{A}_M(\mathbf{u})^{-1}
\boldsymbol{\beta}_M(\mathbf{u}),
\end{equation}
where
\begin{align}
\mathbf{A}_M(\mathbf{u})
&=
\frac{1}{M\sigma_p^2}\mathbf{I}_L
+
\frac{1}{M\sigma^2}
\sum_{m=1}^{M}
\mathbf{F}_L^H
\operatorname{Diag}
\big(
\mathbf{d}(\mathbf{u};\mathbf{y}_m)
\big)
\mathbf{F}_L,\nonumber
\\
\boldsymbol{\beta}_M(\mathbf{u})
&=
\frac{1}{M\sigma_p^2}
\mathbf{F}_L^H
\left(
\mathbf{p}^*\odot\mathbf{y}_p
\right)\nonumber
\\
&\quad+
\frac{1}{M\sigma^2}
\sum_{m=1}^{M}
\mathbf{F}_L^H
\left(
\overline{\mathbf{x}}(\mathbf{u};\mathbf{y}_m)^*
\odot\mathbf{y}_m
\right).
\end{align}
Thus, $\mathbf{A}_M$ and $\boldsymbol{\beta}_M$ are empirical averages
of the posterior moments computed in the E-step. The pilot terms are
$O(M^{-1})$. By \eqref{eq:posterior_envelope} and the uniform strong law,
\begin{equation}
\begin{aligned}
\sup_{\mathbf{u}\in\mathcal{C}}
\left\|
\mathbf{A}_M(\mathbf{u})-\mathbf{A}(\mathbf{u})
\right\|
&=o(1),
\\
\sup_{\mathbf{u}\in\mathcal{C}}
\left\|
\boldsymbol{\beta}_M(\mathbf{u})
-
\boldsymbol{\beta}(\mathbf{u})
\right\|
&=o(1),
\end{aligned}
\qquad
\text{a.s.}
\label{eq:AB_uniform}
\end{equation}
The same uniform convergence holds for
$D\mathbf{A}_M$ and $D\boldsymbol{\beta}_M$. This implies
that the sample EM map and its Jacobian converge uniformly to their
population counterparts on $\mathcal{C}$.

Since $\mathbf{A}(\mathbf{g})=\sigma^{-2}\mathbf{I}_L$, the radius
$r$ can be chosen small enough that $\mathbf{A}(\mathbf{u})$ is
uniformly nonsingular on $\mathcal{C}$. It follows from
\eqref{eq:AB_uniform} that, almost surely,
\begin{equation}
\sup_{\mathbf{u}\in\mathcal{C}}
\left\|
\mathcal{T}_M(\mathbf{u})
-
\mathcal{T}(\mathbf{u})
\right\|
=
o(1),
\label{eq:T_uniform}
\end{equation}
and
\begin{equation}
\sup_{\mathbf{u}\in\mathcal{C}}
\left\|
D\mathcal{T}_M(\mathbf{u})
-
D\mathcal{T}(\mathbf{u})
\right\|_2
=
o(1).
\label{eq:DT_uniform}
\end{equation}
Hence, almost surely for all sufficiently large $M$,
\begin{equation}
\begin{aligned}
\sup_{\mathbf{u}\in\mathcal{C}}
\left\|
\mathcal{T}_M(\mathbf{u})-\mathbf{g}
\right\|
&\le
\kappa r+o(1)
<r,
\\
\sup_{\mathbf{u}\in\mathcal{C}}
\left\|
D\mathcal{T}_M(\mathbf{u})
\right\|_2
&\le
\kappa+o(1)
<1,
\end{aligned}
\label{eq:sample_self_contraction}
\end{equation}
where \eqref{eq:population_contraction} and
\eqref{eq:population_mapping} have been used.

The first inequality in \eqref{eq:sample_self_contraction} ensures that
$\mathcal{T}_M(\mathcal{C})\subset\mathcal{U}\subset\mathcal{C}$.
Moreover, since $\mathcal{C}$ is convex, the second inequality and the
mean-value inequality give, for any
$\mathbf{u},\mathbf{v}\in\mathcal{C}$,
\begin{equation}
\begin{aligned}
\left\|
\mathcal{T}_M(\mathbf{u})
-
\mathcal{T}_M(\mathbf{v})
\right\|
\le
\sup_{\mathbf{w}\in\mathcal{C}}
\left\|
D\mathcal{T}_M(\mathbf{w})
\right\|_2
\left\|
\mathbf{u}-\mathbf{v}
\right\|
<
\left\|
\mathbf{u}-\mathbf{v}
\right\|.
\end{aligned}
\end{equation}
Thus, $\mathcal{T}_M$ is a contraction mapping from $\mathcal{C}$ into
itself. Since $\mathcal{C}$ is closed and hence complete, the Banach
fixed-point theorem guarantees a unique fixed point
$\mathbf{g}_M^\star\in\mathcal{C}$. Because
$\mathcal{T}_M(\mathcal{C})\subset\mathcal{U}$, this fixed point lies
in $\mathcal{U}$. Therefore, every initialization
$\widehat{\mathbf{g}}^{(0)}\in\mathcal{U}$ satisfies
\begin{equation}
\lim_{t\rightarrow\infty}
\widehat{\mathbf{g}}_M^{(t)}
=
\mathbf{g}_M^\star.
\label{eq:sample_fixed_point}
\end{equation}

\subsubsection{Identification With the Local MLE}

It remains to show that the EM fixed point is the local MLE.
Define the normalized observed data log-likelihood as
\begin{equation}
\overline{\ell}_{o,M}(\mathbf{u})
:=
\frac{1}{M}\ell_{o,M}(\mathbf{u}),
\end{equation}
which has the same maximizer as $\ell_{o,M}$. The pilot contribution
is $O(M^{-1})$. By \eqref{eq:likelihood_envelope} and the uniform
strong law,
\begin{equation}
\sup_{\mathbf{u}\in\Theta}
\left\|
\nabla^2\overline{\ell}_{o,M}(\mathbf{u})
-
\nabla^2\overline{\ell}_o(\mathbf{u})
\right\|
=
o(1),
\qquad
\text{a.s.},
\end{equation}
and
\begin{equation}
\left\|
\nabla\overline{\ell}_{o,M}(\mathbf{g})
\right\|
=
o(1),
\qquad
\text{a.s.}
\end{equation}
Together with \eqref{eq:population_concavity}, this implies that,
almost surely for all sufficiently large $M$,
\begin{equation}
\nabla^2\overline{\ell}_{o,M}(\mathbf{u})
\preceq
-\lambda\mathbf{I}_{2L},
\qquad
\mathbf{u}\in\Theta.
\label{eq:sample_concavity}
\end{equation}
Since $\Theta$ is compact and
$\overline{\ell}_{o,M}$ is continuous, it attains a maximum over
$\Theta$. The strong concavity in \eqref{eq:sample_concavity} ensures
that this maximizer is unique. It remains to show that it does not lie
on the boundary. For any $\mathbf{u}\in\partial\Theta$, we have
$\|\mathbf{u}-\mathbf{g}\|=R$, and Taylor's theorem gives
\begin{equation}
\begin{aligned}
\overline{\ell}_{o,M}(\mathbf{u})
-
\overline{\ell}_{o,M}(\mathbf{g})
&\le
R
\left\|
\nabla\overline{\ell}_{o,M}(\mathbf{g})
\right\|
-
\frac{\lambda R^2}{2}
<0
\end{aligned}
\end{equation}
for all sufficiently large $M$. Thus, every boundary point has a lower
likelihood than $\mathbf{g}$, so the unique maximizer must lie in
$\operatorname{int}(\Theta)$.

Since $\mathbf{g}_M^\star$ is an
EM fixed point, using it as the current estimate causes the M-step to
return the same point. Hence,
\begin{equation}
\left.
\nabla_{\mathbf{u}}
\mathcal{L}_M
\left(
\mathbf{u}\mid\mathbf{g}_M^\star
\right)
\right|_{\mathbf{u}=\mathbf{g}_M^\star}
=
\mathbf{0}.
\end{equation}
Here, $\mathcal{L}_M(\mathbf{u}\mid\mathbf{v})$ denotes the E-step
auxiliary function. Moreover, differentiating the observed data log-likelihood gives
\begin{equation}
\begin{aligned}
\nabla\ell_{o,M}(\mathbf{v})
&=
\mathbb{E}_{\mathcal{X}\mid\mathcal{Y},\mathbf{v}}
\left[
\nabla_{\mathbf{v}}
\ell_c(\mathbf{v};\mathcal{D})
\right]
=
\left.
\nabla_{\mathbf{u}}
\mathcal{L}_M(\mathbf{u}\mid\mathbf{v})
\right|_{\mathbf{u}=\mathbf{v}}.
\end{aligned}
\label{eq:em_score_identity}
\end{equation}
Applying \eqref{eq:em_score_identity} at
$\mathbf{v}=\mathbf{g}_M^\star$ yields
\begin{equation}
\nabla\ell_{o,M}(\mathbf{g}_M^\star)
=
\mathbf{0}.
\end{equation}
Since $\mathbf{g}_M^\star\in\mathcal{U}\subset\Theta$ and
$\ell_{o,M}$ is strongly concave on $\Theta$, this stationary point is
the unique maximizer over $\Theta$. Therefore,
\begin{equation}
\mathbf{g}_M^\star
=
\widehat{\mathbf{g}}_M^{\rm ML}.
\end{equation}
Combining this identity with \eqref{eq:sample_fixed_point} proves
\eqref{eq:em_to_mle}.

\section{Proof of Proposition~\ref{prop:mle_consistency}}
\label{app:mle_consistency}

By \eqref{eq:likelihood_envelope} and the uniform strong law,
\begin{equation}
\sup_{\mathbf{u}\in\Theta}
\left|
\overline{\ell}_{o,M}(\mathbf{u})
-
\overline{\ell}_o(\mathbf{u})
\right|
=
o(1),
\qquad
\text{a.s.},
\label{eq:uniform_likelihood}
\end{equation}
where the pilot contribution to
$\overline{\ell}_{o,M}$ is $O(M^{-1})$.

Since
$\nabla\overline{\ell}_o(\mathbf{g})=\mathbf{0}$,
\eqref{eq:population_concavity} and Taylor's theorem give
\begin{equation}
\overline{\ell}_o(\mathbf{g})
-
\overline{\ell}_o(\mathbf{u})
\ge
\lambda
\left\|
\mathbf{u}-\mathbf{g}
\right\|^2,
\qquad
\mathbf{u}\in\Theta.
\label{eq:population_likelihood_gap}
\end{equation}
Thus, for any $0<\epsilon\le R$, every point satisfying
$\|\mathbf{u}-\mathbf{g}\|\ge\epsilon$ has a population likelihood at
least $\lambda\epsilon^2$ below that of $\mathbf{g}$.

By \eqref{eq:uniform_likelihood}, almost surely for all sufficiently
large $M$,
\begin{equation}
\sup_{\mathbf{u}\in\Theta}
\left|
\overline{\ell}_{o,M}(\mathbf{u})
-
\overline{\ell}_o(\mathbf{u})
\right|
<
\frac{\lambda\epsilon^2}{3}.
\end{equation}
Hence, whenever
$\|\mathbf{u}-\mathbf{g}\|\ge\epsilon$,
\begin{equation}
\begin{aligned}
\overline{\ell}_{o,M}(\mathbf{g})
-
\overline{\ell}_{o,M}(\mathbf{u})
&\ge
\lambda\epsilon^2
-
\frac{2\lambda\epsilon^2}{3}
=
\frac{\lambda\epsilon^2}{3}
>0.
\end{aligned}
\end{equation}
Since $\mathbf{g}\in\Theta$, the preceding inequality shows that no
point satisfying $\|\mathbf{u}-\mathbf{g}\|\ge\epsilon$ can maximize
$\overline{\ell}_{o,M}$ over $\Theta$. Hence, almost surely for all
sufficiently large $M$,
\begin{equation}
\left\|
\widehat{\mathbf{g}}_M^{\rm ML}
-
\mathbf{g}
\right\|
<
\epsilon.
\end{equation}
Since this holds for every $\epsilon>0$,
\begin{equation}
\widehat{\mathbf{g}}_M^{\rm ML}
\longrightarrow
\mathbf{g},
\qquad
\text{a.s.},
\end{equation}
which proves \eqref{eq:mle_consistency}.

\section{Proof of Theorem~\ref{thm:em_asymptotic_unbiasedness}}
\label{app:em_asymptotic_unbiasedness}

Condition on the initialization event $\mathcal{E}$. By
Lemma~\ref{prop:em_to_mle},
$\widehat{\mathbf{g}}_M^{(\infty)}$ agrees almost surely with
$\widehat{\mathbf{g}}_M^{\rm ML}$ for all sufficiently large $M$.
Proposition~\ref{prop:mle_consistency} therefore gives
\begin{equation}
\widehat{\mathbf{g}}_M^{(\infty)}
\longrightarrow
\mathbf{g},
\qquad
\text{a.s. on }\mathcal{E}.
\label{eq:em_almost_sure_consistency}
\end{equation}

By the definition of the completed EM output,
$\widehat{\mathbf{g}}_M^{(\infty)}\in\Theta$ on $\mathcal{E}$ for
every $M$. Since
\begin{equation}
\Theta
=
\left\{
\mathbf{u}\in\mathbb{C}^L:
\|\mathbf{u}-\mathbf{g}\|\le R
\right\},
\end{equation}
we have
\begin{equation}
\left\|
\widehat{\mathbf{g}}_M^{(\infty)}
-
\mathbf{g}
\right\|
\le
R.
\end{equation}
Thus, the estimation error is bounded by a fixed integrable constant.
Applying conditional dominated convergence to
\eqref{eq:em_almost_sure_consistency} gives
\begin{equation}
\mathbb{E}
\left[
\left\|
\widehat{\mathbf{g}}_M^{(\infty)}
-
\mathbf{g}
\right\|
\middle|
\mathcal{E}
\right]
\longrightarrow
0.
\end{equation}
Consequently,
\begin{equation}
\begin{aligned}
\left\|
\mathbb{E}
\left[
\widehat{\mathbf{g}}_M^{(\infty)}
\mid
\mathcal{E}
\right]
-
\mathbf{g}
\right\|
&\le
\mathbb{E}
\left[
\left\|
\widehat{\mathbf{g}}_M^{(\infty)}
-
\mathbf{g}
\right\|
\middle|
\mathcal{E}
\right]
\longrightarrow
0,
\end{aligned}
\end{equation}
which proves \eqref{eq:em_asymptotic_unbiasedness}.

\section{Proof of Theorem~\ref{thm:em_mse}}
\label{app:em_mse}
Throughout this proof, all probabilities and expectations are
conditioned on $\mathcal{E}$. Since $\mathcal{E}$ depends only on the
pilot observation, the payload observations remain i.i.d. after
conditioning. The proof compares the converged EM estimator with the
local MLE. We first show that, with probability
$1-O(M^{-2})$, both estimators are well defined locally and coincide.
We then derive the $1/M$ MSE expansion of the local MLE from
its asymptotic normality. Finally, since the remaining event has
probability $O(M^{-2})$, it contributes only $o(M^{-1})$ to the MSE of the EM estimator, so EM has the same first-order MSE as ML.

\subsubsection{Finite-Sample Stability}

Appendix~\ref{app:em_to_mle} establishes the desired local EM and ML
properties for all sufficiently large $M$ almost surely. For the MSE
analysis, we also need to bound their finite-$M$ failure probabilities.

Define the EM stability event
\begin{equation}
\mathcal{G}_M
:=
\left\{
\begin{aligned}
\sup_{\mathbf{u}\in\mathcal{C}}
\left\|
\mathcal{T}_M(\mathbf{u})
-
\mathcal{T}(\mathbf{u})
\right\|
&\le
\frac{(1-\kappa)r}{2},
\\
\sup_{\mathbf{u}\in\mathcal{C}}
\left\|
D\mathcal{T}_M(\mathbf{u})
-
D\mathcal{T}(\mathbf{u})
\right\|_2
&\le
\frac{1-\kappa}{2}
\end{aligned}
\right\}.
\label{eq:em_stability_event}
\end{equation}
Using \eqref{eq:population_contraction} and
\eqref{eq:population_mapping}, on $\mathcal{G}_M$,
\begin{equation}
\begin{aligned}
\sup_{\mathbf{u}\in\mathcal{C}}
\left\|
\mathcal{T}_M(\mathbf{u})-\mathbf{g}
\right\|
&\le
\frac{1+\kappa}{2}r
<r,
\\
\sup_{\mathbf{u}\in\mathcal{C}}
\left\|
D\mathcal{T}_M(\mathbf{u})
\right\|_2
&\le
\frac{1+\kappa}{2}
<1.
\end{aligned}
\end{equation}
Thus, $\mathcal{T}_M$ maps $\mathcal{C}$ into itself and is a
contraction there.

Similarly, define the ML stability event
\begin{equation}
\mathcal{H}_M
:=
\left\{
\begin{aligned}
\sup_{\mathbf{u}\in\Theta}
\left\|
\nabla^2\overline{\ell}_{o,M}(\mathbf{u})
-
\nabla^2\overline{\ell}_o(\mathbf{u})
\right\|
&\le
\lambda,
\\
\left\|
\nabla\overline{\ell}_{o,M}(\mathbf{g})
\right\|
&\le
\frac{\lambda R}{4}
\end{aligned}
\right\}.
\label{eq:ml_stability_event}
\end{equation}
By \eqref{eq:population_concavity}, the first condition implies
\begin{equation}
\nabla^2\overline{\ell}_{o,M}(\mathbf{u})
\preceq
-\lambda\mathbf{I}_{2L},
\qquad
\mathbf{u}\in\Theta.
\end{equation}
Moreover, for every $\mathbf{u}\in\partial\Theta$, the second condition
and Taylor's theorem give
\begin{equation}
\overline{\ell}_{o,M}(\mathbf{u})
-
\overline{\ell}_{o,M}(\mathbf{g})
\le
\frac{\lambda R^2}{4}
-
\frac{\lambda R^2}{2}
<0.
\end{equation}
Hence, on $\mathcal{H}_M$, the sample likelihood is strongly concave
on $\Theta$ and has a unique maximizer in
$\operatorname{int}(\Theta)$.

It remains to bound the probabilities of
$\mathcal{G}_M^c$ and $\mathcal{H}_M^c$. After removing the
$O(M^{-1})$ pilot terms, every real scalar entry of
$\mathbf{A}_M-\mathbf{A}$,
$\boldsymbol{\beta}_M-\boldsymbol{\beta}$, their derivatives, the
sample score, and the sample Hessian is an average of $M$ centered
payload contributions. For any one of these entries, denote its
centered single-symbol contribution by $Z_m(\mathbf{u})$. For each
fixed $\mathbf{u}$, the variables
$\{Z_m(\mathbf{u})\}_{m=1}^M$ are i.i.d., zero mean, and have finite
fourth moments. Let $\overline{Z}_M(\mathbf{u}): = \frac{1}{M}
\sum_{m=1}^{M}
Z_m(\mathbf{u})$. Then,
\begin{equation}
\mathbb{E}
\left[
\left|
\overline{Z}_M(\mathbf{u})
\right|^4
\right]
=
O(M^{-2}).
\label{eq:empirical_fourth_moment}
\end{equation}
Markov's inequality then gives, for every $c>0$,
\begin{equation}
\begin{aligned}
\Pr
\left(
\left|
\overline{Z}_M(\mathbf{u})
\right|
>c
\right)
\le
\frac{1}{c^4}
\mathbb{E}
\left[
\left|
\overline{Z}_M(\mathbf{u})
\right|^4
\right]
=
O(M^{-2}).
\end{aligned}
\label{eq:pointwise_deviation_probability}
\end{equation}

This bound initially holds at a fixed $\mathbf{u}$. The envelope
bounds in \eqref{eq:likelihood_envelope} and
\eqref{eq:posterior_envelope} provide i.i.d. random variables $L_m$
such that
\begin{equation}
\left|
Z_m(\mathbf{u})-Z_m(\mathbf{v})
\right|
\le
L_m
\|\mathbf{u}-\mathbf{v}\|,
\qquad
\mathbb{E}[L_m^4]<\infty.
\end{equation}
Thus, the empirical average has Lipschitz factor $\overline{L}_M
:=
\frac{1}{M}
\sum_{m=1}^{M}
L_m.$ For any fixed $L_0>\mathbb{E}[L_1]$, the same fourth-moment argument
gives
\begin{equation}
\begin{aligned}
\Pr(\overline{L}_M>L_0)
&\le
\frac{
\mathbb{E}
\left[
\left|
\overline{L}_M-\mathbb{E}[L_1]
\right|^4
\right]
}{
\left(
L_0-\mathbb{E}[L_1]
\right)^4
}
=
O(M^{-2}).
\end{aligned}
\label{eq:lipschitz_factor_probability}
\end{equation}

Cover $\mathcal{C}$ by finitely many points
$\{\mathbf{u}_1,\ldots,\mathbf{u}_{N_\delta}\}$ such that every
$\mathbf{u}\in\mathcal{C}$ lies within $\delta$ of one covering point,
where $\delta<c/(2L_0)$. On
$\{\overline{L}_M\le L_0\}$, a deviation larger than $c$ somewhere in
$\mathcal{C}$ implies a deviation larger than $c/2$ at one of the
covering points. Hence,
\begin{equation}
\begin{aligned}
&
\Pr
\left(
\sup_{\mathbf{u}\in\mathcal{C}}
\left|
\overline{Z}_M(\mathbf{u})
\right|
>c
\right)
\\
&\le
\sum_{i=1}^{N_\delta}
\Pr
\left(
\left|
\overline{Z}_M(\mathbf{u}_i)
\right|
>
\frac{c}{2}
\right)
+
\Pr(\overline{L}_M>L_0)=
O(M^{-2}).
\end{aligned}
\label{eq:uniform_deviation_probability}
\end{equation}

Since the relevant matrices, vectors, and derivative arrays have
finitely many real scalar entries, a finite union bound preserves the
same order. Consequently, for every fixed $c>0$,
\begin{equation}
\begin{aligned}
\Pr
\left(
\sup_{\mathbf{u}\in\mathcal{C}}
\left\|
\mathbf{A}_M(\mathbf{u})-\mathbf{A}(\mathbf{u})
\right\|>c
\right)
&=O(M^{-2}),
\\
\Pr
\left(
\sup_{\mathbf{u}\in\mathcal{C}}
\left\|
\boldsymbol{\beta}_M(\mathbf{u})
-\boldsymbol{\beta}(\mathbf{u})
\right\|>c
\right)
&=O(M^{-2}),
\end{aligned}
\label{eq:AB_probability}
\end{equation}
and
\begin{equation}
\begin{aligned}
\Pr
\left(
\sup_{\mathbf{u}\in\mathcal{C}}
\left\|
D\mathbf{A}_M(\mathbf{u})
-D\mathbf{A}(\mathbf{u})
\right\|>c
\right)
&=O(M^{-2}),
\\
\Pr
\left(
\sup_{\mathbf{u}\in\mathcal{C}}
\left\|
D\boldsymbol{\beta}_M(\mathbf{u})
-D\boldsymbol{\beta}(\mathbf{u})
\right\|>c
\right)
&=O(M^{-2}).
\end{aligned}
\label{eq:DAB_probability}
\end{equation}

Since $\mathbf{A}(\mathbf{u})$ is uniformly nonsingular on
$\mathcal{C}$, sufficiently small deviations in
\eqref{eq:AB_probability} and \eqref{eq:DAB_probability} imply both
conditions defining $\mathcal{G}_M$. A finite union bound therefore
gives
\begin{equation}
\Pr(\mathcal{G}_M^c)
=
O(M^{-2}).
\end{equation}

Applying the same argument over $\Theta$ to the sample score and
Hessian gives
\begin{equation}
\begin{aligned}
\Pr
\left(
\sup_{\mathbf{u}\in\Theta}
\left\|
\nabla^2\overline{\ell}_{o,M}(\mathbf{u})
-
\nabla^2\overline{\ell}_o(\mathbf{u})
\right\|>\lambda
\right)
&=O(M^{-2}),
\\
\Pr
\left(
\left\|
\nabla\overline{\ell}_{o,M}(\mathbf{g})
\right\|>
\frac{\lambda R}{4}
\right)
&=O(M^{-2}).
\end{aligned}
\end{equation}
These are precisely the two failure conditions in
$\mathcal{H}_M^c$, so
\begin{equation}
\Pr(\mathcal{H}_M^c)
=
O(M^{-2}).
\end{equation}
Combining the two bounds yields
\begin{equation}
\Pr(\mathcal{G}_M^c)
+
\Pr(\mathcal{H}_M^c)
=
O(M^{-2}).
\label{eq:stability_event_rate}
\end{equation}
The normalized pilot terms are $O(M^{-1})$ and do not affect these
probability orders.

\subsubsection{MSE of the Local MLE}

Let
\begin{equation}
\widehat{\boldsymbol{\theta}}_M^{\rm ML}
:=
\boldsymbol{\theta}
\left(
\widehat{\mathbf{g}}_M^{\rm ML}
\right),
\qquad
\boldsymbol{\theta}^{\star}
:=
\boldsymbol{\theta}(\mathbf{g}).
\end{equation}
The smoothness established in Appendix~\ref{app:em_to_mle},
the consistency in Proposition~\ref{prop:mle_consistency}, and
$\mathbf{J}_o\succ\mathbf{0}$ give the standard ML limit
\cite[Theorem~3.1]{RednerWalker1984}:
\begin{equation}
\sqrt{M}
\left(
\widehat{\boldsymbol{\theta}}_M^{\rm ML}
-
\boldsymbol{\theta}^{\star}
\right)
\xrightarrow{d}
\mathcal{N}
\left(
\mathbf{0},
\mathbf{J}_o^{-1}
\right).
\label{eq:mle_asymptotic_normality}
\end{equation}
The fixed pilot contributes only $O(1)$ to the unnormalized score and
Hessian and therefore does not affect this limit.

Convergence in distribution alone does not guarantee convergence of
the MSE. Define the normalized payload score
\begin{equation}
\mathbf{s}_M
:=
\frac{1}{\sqrt{M}}
\sum_{m=1}^{M}
\mathbf{s}
\left(
\mathbf{y}_m;
\boldsymbol{\theta}^{\star}
\right).
\end{equation}
Since each payload score has zero mean and a finite fourth moment, we have $\sup_M
\mathbb{E}
\left[
\|\mathbf{s}_M\|^4
\right]
<
\infty$. On $\mathcal{H}_M$, the local MLE is an interior stationary
point. Applying Taylor's theorem to the likelihood score between
$\boldsymbol{\theta}^{\star}$ and
$\widehat{\boldsymbol{\theta}}_M^{\rm ML}$, together with the Hessian
bound in \eqref{eq:sample_concavity}, gives
\begin{equation}
\begin{aligned}
\sqrt{M}
\left\|
\widehat{\boldsymbol{\theta}}_M^{\rm ML}
-
\boldsymbol{\theta}^{\star}
\right\|
\le
\frac{1}{\lambda}
\Bigg\|
\mathbf{s}_M
+
\frac{1}{\sqrt{M}}
\left.
\nabla_{\boldsymbol{\theta}}
\log
p_{\boldsymbol{\theta}}
(\mathbf{y}_p\mid\mathbf{p})
\right|_{\boldsymbol{\theta}
=\boldsymbol{\theta}^{\star}}
\Bigg\|.
\end{aligned}
\label{eq:scaled_mle_bound}
\end{equation}
The pilot score has a finite fourth moment, so the fourth moment of the scaled
ML error on $\mathcal{H}_M$ is also bounded. On $\mathcal{H}_M^c$, both
$\widehat{\mathbf{g}}_M^{\rm ML}$ and $\mathbf{g}$ lie in $\Theta$,
and hence
\begin{equation}
\left\|
\widehat{\boldsymbol{\theta}}_M^{\rm ML}
-
\boldsymbol{\theta}^{\star}
\right\|
\le
R.
\end{equation}
Splitting the expectation over $\mathcal{H}_M$ and
$\mathcal{H}_M^c$, and using
$M^2R^4\Pr(\mathcal{H}_M^c)=O(1)$, gives
\begin{equation}
\sup_M
\mathbb{E}
\left[
\left\|
\sqrt{M}
\left(
\widehat{\boldsymbol{\theta}}_M^{\rm ML}
-
\boldsymbol{\theta}^{\star}
\right)
\right\|^4
\right]
<
\infty.
\label{eq:mle_fourth_moment_bound}
\end{equation}
Therefore,
$M|\widehat g_{M,k}^{\rm ML}-g_k|^2$ is uniformly integrable.
Combining this property with
\eqref{eq:mle_asymptotic_normality} gives, for each fixed
$k=0,\ldots,L-1$,
\begin{equation}
\begin{aligned}
&
\mathbb{E}
\left[
\left|
\widehat g_{M,k}^{\rm ML}
-
g_k
\right|^2
\right]
\\
&\quad=
\frac{
[\mathbf{J}_o^{-1}]_{k+1,k+1}
+
[\mathbf{J}_o^{-1}]_{L+k+1,L+k+1}
}{M}
+
o(M^{-1}).
\end{aligned}
\label{eq:mle_mse}
\end{equation}

\subsubsection{Transfer from ML to EM}

On $\mathcal{G}_M\cap\mathcal{H}_M$, the converged EM estimator equals
the local MLE. Any difference between them can therefore occur
only on the complementary event. Since both estimators lie in
$\Theta$,
\begin{equation}
\left\|
\widehat{\mathbf{g}}_M^{(\infty)}
-
\widehat{\mathbf{g}}_M^{\rm ML}
\right\|
\le
2R.
\end{equation}
Using \eqref{eq:stability_event_rate}, we obtain
\begin{equation}
\begin{aligned}
&
\mathbb{E}
\left[
\left\|
\widehat{\mathbf{g}}_M^{(\infty)}
-
\widehat{\mathbf{g}}_M^{\rm ML}
\right\|^2
\right]
\le
4R^2
\Pr
\left(
\mathcal{G}_M^c
\cup
\mathcal{H}_M^c
\right)
=
O(M^{-2}).
\end{aligned}
\label{eq:em_ml_transfer}
\end{equation}

By the Cauchy--Schwarz inequality, \eqref{eq:mle_mse}, and
\eqref{eq:em_ml_transfer},
\begin{equation}
\begin{aligned}
&
\left|
\mathbb{E}
\left[
\left|
\widehat g_{M,k}^{(\infty)}
-
g_k
\right|^2
\right]
-
\mathbb{E}
\left[
\left|
\widehat g_{M,k}^{\rm ML}
-
g_k
\right|^2
\right]
\right|
\\
&\le
2
\left(
\mathbb{E}
\left[
\left|
\widehat g_{M,k}^{\rm ML}
-
g_k
\right|^2
\right]
\right)^{1/2}
\left(
\mathbb{E}
\left[
\left|
\widehat g_{M,k}^{(\infty)}
-
\widehat g_{M,k}^{\rm ML}
\right|^2
\right]
\right)^{1/2}
\\
&+
\mathbb{E}
\left[
\left|
\widehat g_{M,k}^{(\infty)}
-
\widehat g_{M,k}^{\rm ML}
\right|^2
\right]
=
O(M^{-3/2})
+
O(M^{-2})
=
o(M^{-1}).
\end{aligned}
\end{equation}
Thus, the EM and MLEs have the same first-order MSE.
Substituting \eqref{eq:mle_mse} proves
\eqref{eq:em_mse_result}.

\balance


\begin{thebibliography}{99}

\bibitem{ref1}
Z. Wei \emph{et al.}, ``Integrated sensing and communication signals toward 5G-A and 6G: A survey,'' \emph{IEEE Internet Things J.}, vol. 10, no. 13, pp. 11068--11092, 2023.

\bibitem{ref2}
F. Liu, Y. Cui, C. Masouros, J. Xu, T. X. Han, Y. C. Eldar, and S. Buzzi, ``Integrated sensing and communications: Toward dual-functional wireless networks for 6G and beyond,'' \emph{IEEE J. Sel. Areas Commun.}, vol. 40, no. 6, pp. 1728--1767, Jun. 2022.

\bibitem{ref3}
B. Paul, A. R. Chiriyath, and D. W. Bliss, ``Survey of RF communications and sensing convergence research,'' \emph{IEEE Access}, vol. 5, pp. 252--270, 2017.

\bibitem{ref4}
J. A. Zhang, F. Liu, C. Masouros, R. W. Heath Jr., Z. Feng, L. Zheng, and A. Petropulu, ``An overview of signal processing techniques for joint communication and radar sensing,'' \emph{IEEE J. Sel. Topics Signal Process.}, vol. 15, no. 6, pp. 1295--1315, Nov. 2021.

\bibitem{ref5}
F. Liu, C. Masouros, A. P. Petropulu, H. Griffiths, and L. Hanzo, ``Joint radar and communication design: Applications, state-of-the-art, and the road ahead,'' \emph{IEEE Trans. Commun.}, vol. 68, no. 6, pp. 3834--3862, 2020.

\bibitem{ref6}
S. Lu \emph{et al.}, ``Integrated sensing and communications: Recent advances and ten open challenges,'' \emph{IEEE Internet Things J.}, vol. 11, no. 11, pp. 19094--19120, 2024.

\bibitem{ref7}
F. Liu, L. Zheng, Y. Cui, C. Masouros, A. P. Petropulu, H. Griffiths, and Y. C. Eldar, ``Seventy years of radar and communications: The road from separation to integration,'' \emph{IEEE Signal Process. Mag.}, vol. 40, no. 5, pp. 106--121, 2023.

\bibitem{ref8}
W. Zhou, R. Zhang, G. Chen, and W. Wu, ``Integrated sensing and communication waveform design: A survey,'' \emph{IEEE Open J. Commun. Soc.}, vol. 3, pp. 1930--1949, 2022.

\bibitem{ref9}
S. P. Chepuri, N. Shlezinger, F. Liu, G. C. Alexandropoulos, S. Buzzi, and Y. C. Eldar, ``Integrated sensing and communications with reconfigurable intelligent surfaces: From signal modeling to processing,'' \emph{IEEE Signal Process. Mag.}, vol. 40, no. 6, pp. 41--62, 2023.

\bibitem{ref10}
C. Sturm, T. Zwick, and W. Wiesbeck, ``An OFDM system concept for joint radar and communications operations,'' in \emph{Proc. IEEE 69th Veh. Technol. Conf. (VTC Spring)}, Barcelona, Spain, 2009, pp. 1--5.

\bibitem{ref11}
C. Sturm and W. Wiesbeck, ``Waveform design and signal processing aspects for fusion of wireless communications and radar sensing,'' \emph{Proc. IEEE}, vol. 99, no. 7, pp. 1236--1259, Jul. 2011.

\bibitem{ref12}
M. Braun, ``OFDM radar algorithms in mobile communication networks,'' Ph.D. dissertation, Karlsruhe Institute of Technology, Karlsruhe, Germany, 2014.

\bibitem{ref13}
Y. Liu, G. Liao, Y. Chen, J. Xu, and Y. Yin, ``Super-resolution range and velocity estimations with OFDM integrated radar and communications waveform,'' \emph{IEEE Trans. Veh. Technol.}, vol. 69, no. 10, pp. 11659--11672, 2020.

\bibitem{ref14}
Y. Liu, G. Liao, Z. Yang, and J. Xu, ``Design of integrated radar and communication system based on MIMO-OFDM waveform,'' \emph{J. Syst. Eng. Electron.}, vol. 28, no. 4, pp. 669--680, 2017.

\bibitem{LiuIceberg2025}
F. Liu, Y. Xiong, S. Lu, S. Li, W. Yuan, C. Masouros, S. Jin, and G. Caire, ``Uncovering the iceberg in the sea: Fundamentals of pulse shaping and modulation design for random ISAC signals,'' \emph{IEEE Trans. Signal Process.}, vol. 73, pp. 2511--2526, 2025.

\bibitem{Lu2024RandomPrecoding}
S. Lu, F. Liu, F. Dong, Y. Xiong, J. Xu, Y.-F. Liu, and S. Jin,
``Random ISAC signals deserve dedicated precoding,''
\emph{IEEE Trans. Signal Process.},
vol. 72, pp. 3453--3469, 2024.

\bibitem{ref21}
D. Brunner \emph{et al.}, ``Bistatic OFDM-based ISAC with over-the-air synchronization: System concept and performance analysis,'' \emph{IEEE Trans. Microw. Theory Techn.}, vol. 73, no. 5, pp. 3016--3029, May 2025.

\bibitem{ref25}
M. Henninger, L. Giroto, A. Elkelesh, and S. Mandelli, ``Hybrid resource allocation scheme for bistatic ISAC with data channels,'' 2026, \emph{arXiv:2601.11110}.

\bibitem{ref24}
M. F. Keskin, S. Mura, M. Mizmizi, D. Tagliaferri, and H. Wymeersch, ``Bridging the gap via data-aided sensing: Can bistatic ISAC converge to genie performance?,'' in \emph{Proc. IEEE Radar Conf. (RadarConf25)}, 2025, pp. 514--519.

\bibitem{KeskinPayload2026}
M. F. Keskin, S. Mura, M. Mizmizi, D. Tagliaferri, F. Liu, and H. Wymeersch, ``Bistatic ISAC under data payload uncertainty,'' 2026, \emph{arXiv:2610.05527}.

\bibitem{ref15}
T. Riihonen and K. E. Kolodziej, ``Full-duplex ISAC,'' in \emph{Integrated Sensing and Communications}, F. Liu, C. Masouros, and Y. C. Eldar, Eds. Singapore: Springer, 2023, pp. 537--565.

\bibitem{ref16}
A. Tang, X. Wang, and J. A. Zhang, ``Interference management for full-duplex ISAC in B5G/6G networks: Architectures, challenges, and solutions,'' \emph{IEEE Commun. Mag.}, vol. 62, no. 9, pp. 20--26, 2024.

\bibitem{ref17}
Z. Zhang, X. Chai, K. Long, A. V. Vasilakos, and L. Hanzo, ``Full duplex techniques for 5G networks: Self-interference cancellation, protocol design, and relay selection,'' \emph{IEEE Commun. Mag.}, vol. 53, no. 5, pp. 128--137, 2015.

\bibitem{ref19}
J. Zhou, T. H. Chuang, T. Dinc, and H. Krishnaswamy, ``Integrated wideband self-interference cancellation in the RF domain for FDD and full-duplex wireless,'' \emph{IEEE J. Solid-State Circuits}, vol. 50, no. 12, pp. 3015--3031, 2015.

\bibitem{ref20}
M. Chung, M. S. Sim, J. Kim, D. K. Kim, and C.-B. Chae, ``Prototyping real-time full duplex radios,'' \emph{IEEE Commun. Mag.}, vol. 53, no. 9, pp. 56--63, 2015.

\bibitem{ref22}
L. G. de Oliveira \emph{et al.}, ``Bistatic OFDM-based joint radar-communication: Synchronization, data communication and sensing,'' in \emph{Proc. 20th Eur. Radar Conf. (EuRAD)}, Berlin, Germany, 2023, pp. 359--362.

\bibitem{Valiulahi2026LANM}
I. Valiulahi, C. Masouros, and A. P. Petropulu, ``ISAC super-resolution receiver via lifted atomic norm minimization,'' \emph{IEEE Trans. Commun.}, vol. 74, pp. 5184--5198, 2026.

\bibitem{Xie2025}
L. Xie, F. Liu, S. Song, and S. Jin,
``Bistatic target detection by exploiting both deterministic pilots
and unknown random data payloads,''
\emph{IEEE Trans. Wireless Commun.},
2026, \emph{early access}.

\bibitem{Hatami2026}
M. Hatami, N. T. Nguyen, and M. Juntti,
``Beamforming and filter design for bistatic ISAC under known
and unknown transmit symbols,''
to appear in \emph{Proc. IEEE Int. Workshop Signal Process.
Artif. Intell. Wireless Commun. (SPAWC)}, 2026.

\bibitem{ref29}
A. P. Dempster, N. M. Laird, and D. B. Rubin, ``Maximum likelihood from incomplete data via the EM algorithm,'' \emph{J. Roy. Stat. Soc. Ser. B}, vol. 39, no. 1, pp. 1--38, 1977.

\bibitem{ref30}
T. K. Moon, ``The expectation-maximization algorithm,'' \emph{IEEE Signal Process. Mag.}, vol. 13, no. 6, pp. 47--60, Nov. 1996.

\bibitem{Ladaycia2019}
A. Ladaycia, A. Belouchrani, K. Abed-Meraim, and A. Mokraoui, ``Semi-blind MIMO-OFDM channel estimation using expectation maximisation like techniques,'' \emph{IET Commun.}, vol. 13, no. 20, pp. 3452--3462, 2019.

\bibitem{RednerWalker1984}
R. A. Redner and H. F. Walker, ``Mixture densities, maximum likelihood and the EM algorithm,'' \emph{SIAM Rev.}, vol. 26, no. 2, pp. 195--239, Apr. 1984.

\bibitem{Ma2004EM}
X. Ma, H. Kobayashi, and S. C. Schwartz,
``EM-based channel estimation algorithms for OFDM,''
\emph{EURASIP J. Appl. Signal Process.}, vol. 2004, no. 10,
pp. 1460--1477, 2004.

\bibitem{Balakrishnan2017EM}
S. Balakrishnan, M. J. Wainwright, and B. Yu,
``Statistical guarantees for the EM algorithm: From population to
sample-based analysis,''
\emph{Ann. Statist.}, vol. 45, no. 1,
pp. 77--120, Feb. 2017.

\end{thebibliography}
\end{document}